\documentclass{article}

\usepackage{lineno,hyperref}
\modulolinenumbers[5]
\usepackage{gensymb}
\usepackage{amsmath}
\usepackage[letterpaper, margin=1in]{geometry}
\usepackage{graphicx}
\usepackage{mhchem}
\usepackage{upgreek}
\usepackage{braket}
\usepackage{float}
\usepackage{xcolor}
\usepackage{ulem}

\graphicspath{{numbered_figures/}}

\newcommand\sig{\textit{S}}
\newcommand\noise{\textit{N}}
\newcommand\snr{\textit{SNR}}
\newcommand\NA{\textrm{NA}}

\newcommand\nt{\textup{\scriptsize{0}}}

\newcommand\el{\textup{\scriptsize{el}}}
\newcommand\inel{\textup{\scriptsize{inel}}}
\newcommand\elin{\textup{\scriptsize{el,in}}}
\newcommand\inelin{\textup{\scriptsize{inel,in}}}
\newcommand\out{\textup{\scriptsize{out}}}
\newcommand\PI{\textup{\scriptsize{abs}}}
\newcommand\noscat{\textup{\scriptsize{noscat}}}
\newcommand\sel{\textup{\scriptsize{1el}}}
\newcommand\elpl{\textup{\scriptsize{el,plural}}}
\newcommand\innoinel{\textup{\scriptsize{in,noinel}}}

\newcommand\signali{\textup{\scriptsize{signal,\textit i}}}
\newcommand\signalf{\textup{\scriptsize{signal,\textit f}}}
\newcommand\signalb{\textup{\scriptsize{signal,\textit b}}}

\newcommand\pcf{\textup{\scriptsize{noise},\textit f}}
\newcommand\pcb{\textup{\scriptsize{noise},\textit b}}

\newcommand\elb{\textup{\scriptsize{el,\textit b}}}
\newcommand\inelb{\textup{\scriptsize{inel,\textit b}}}
\newcommand\elinb{\textup{\scriptsize{el,in,\textit b}}}
\newcommand\inelinb{\textup{\scriptsize{inel,in,\textit b}}}
\newcommand\outb{\textup{\scriptsize{out,\textit b}}}
\newcommand\PIb{\textup{\scriptsize{abs,\textit b}}}
\newcommand\noscatb{\textup{\scriptsize{noscat,\textit b}}}
\newcommand\selb{\textup{\scriptsize{1el,\textit b}}}
\newcommand\elplb{\textup{\scriptsize{el,plural,\textit b}}}
\newcommand\innoinelb{\textup{\scriptsize{in,noinel,\textit b}}}

\newcommand\inb{\textup{\scriptsize{in,\textit b}}}

\newcommand\elf{\textup{\scriptsize{el,\textit f}}}
\newcommand\inelf{\textup{\scriptsize{inel,\textit f}}}
\newcommand\elinf{\textup{\scriptsize{el,in,\textit f}}}
\newcommand\inelinf{\textup{\scriptsize{inel,in,\textit f}}}
\newcommand\outf{\textup{\scriptsize{out,\textit f}}}
\newcommand\outff{\textup{\scriptsize{out/f,\textit f}}}
\newcommand\PIf{\textup{\scriptsize{abs,\textit f}}}
\newcommand\PIff{\textup{\scriptsize{abs/f,\textit f}}}
\newcommand\noscatf{\textup{\scriptsize{noscat,\textit f}}}
\newcommand\self{\textup{\scriptsize{1el,\textit f}}}
\newcommand\selff{\textup{\scriptsize{1el/f,\textit f}}}
\newcommand\elplf{\textup{\scriptsize{el,plural,\textit f}}}
\newcommand\innoinelf{\textup{\scriptsize{in,noinel,\textit f}}}

\newcommand\inff{\textup{\scriptsize{in,\textit f}}}

\newcommand\pci{\textup{\scriptsize{noise},\textit i}}

\newcommand\ineli{\textup{\scriptsize{inel,\textit i}}}

\newcommand\PIi{\textup{\scriptsize{abs,\textit i}}}

\newcommand\elpli{\textup{\scriptsize{el,plural,\textit i}}}

\newcommand\micron{$\upmu$m}
\newcommand\zpcf{\textup{\scriptsize{\textit f,zpc}}}
\newcommand\zpcb{\textup{\scriptsize{\textit b,zpc}}}

\newcommand\absf{\textup{\scriptsize{\textit f,abs}}}
\newcommand\absb{\textup{\scriptsize{\textit b,abs}}}

\begin{document}

\date{}

\title{Relative merits and limiting factors for x-ray and electron
  microscopy of thick, hydrated organic materials \textcolor{blue}{(2020 revised version)}}
\maketitle
\vspace{-0.5in}
%% Group authors per affiliation:
\begin{center}
	\author{Ming Du$^1$, and Chris Jacobsen$^{2,3,4,*}$}
\end{center}
\scriptsize{$^{1}$Department of Materials Science and Engineering,
  Northwestern University, 2145 Sheridan Road, Evanston IL 60208, USA}\\
\scriptsize{$^{2}$Advanced Photon Source, Argonne National Laboratory,
  9700 South Cass Avenue, Argonne IL 60439, USA}\\
\scriptsize{$^{3}$Department of Physics \&{} Astronomy,
  Northwestern University, 2145 Sheridan Road, Evanston IL 60208, USA}\\
\scriptsize{$^{4}$Chemistry of Life Processes Institute, Northwestern
  University, 2170 Campus Drive, Evanston IL 60208, USA}\\
\scriptsize{$^{*}$Corresponding author; Email: cjacobsen@anl.gov}
\normalsize

\begin{abstract}
  Electron and x-ray microscopes allow one to image the entire,
  unlabeled structure of hydrated materials at a resolution well
  beyond what visible light microscopes can achieve.  However, both
  approaches involve ionizing radiation, so that radiation damage must
  be considered as one of the limits to imaging.  Drawing upon earlier
  work, we describe here a unified approach to estimating the image
  contrast (and thus the required exposure and corresponding radiation
  dose) in both x-ray and electron microscopy.  This approach accounts
  for factors such as plural and inelastic scattering, and (in
  electron microscopy) the use of energy filters to obtain so-called
  ``zero loss'' images.  As expected, it shows that electron
  microscopy offers lower dose for specimens thinner than about 1
  \micron\ (such as for studies of macromolecules, viruses, bacteria and
  archaebacteria, and thin sectioned material), while x-ray microscopy
  offers superior characteristics for imaging thicker specimen such as
  whole eukaryotic cells, thick-sectioned tissues, and organs.  The
  required radiation dose scales strongly as a function of the desired
  spatial resolution, allowing one to understand the limits of live
  and frozen hydrated specimen imaging.  Finally, we consider the
  factors limiting x-ray microscopy of thicker materials, suggesting
  that specimens as thick as a whole mouse brain can be imaged with
  x-ray microscopes without significant image degradation should
  appropriate image reconstruction methods be identified.
  \textcolor{blue}{The as-published article [\emph{Ultramicroscopy}
    \textbf{184}, 293--309 (2018); doi:10.1016/j.ultramic.2017.10.003]
    had some minor mistakes that we correct here, with all changes
    from the as-published article shown in blue.}
\end{abstract}

\begin{center}
\end{center}

\section{Introduction}
\label{sec:introduction}

Soft materials are often wet materials. This applies to polymer gel
absorbants, biofilms in soils, and in particular to biological
specimens.  While tremendous advances have been made in
superresolution light microscopy for the study of added fluorophores
or genetically-encoded fluorescent regions in proteins, it is still
desirable to image the entirety of a specimen using its intrinsic
contrast to provide overall structural context and for this the
classical resolution limit of $\delta=0.61\lambda/\NA$
applies.  Given the numerical aperture of wet or oil immersion lenses,
this means that it is difficult to see sub-100 nm intrinsic detail in
soft materials using visible light microscopes, even when using methods such as structured
illumination microscopy \cite{gustafsson_jmic_2000}.

For higher resolution, one must use radiation with a shorter
wavelength $\lambda$, which is provided in both electron and x-ray
microscopes.  However, 
these shorter wavelength probes come at
a cost in loss of convenience and, more fundamentally, a cost
in terms of radiation damage caused by the use of ionizing radiation.
For simple linear effects in molecules, Burton proposed
\cite{burton_fd_1952} a measure called the $G$ value which is the number
of irreversible molecular damage events caused per 100 eV of absorbed
energy. While $G$
values for organic molecules span a wide range (such as 0.7 for
styrene and 12 for methyl methacrylate at low dose rates
\cite{chapiro_1962}), a geometrical mean of 2.9 implies that a
molecular bond is broken for every 35 eV of ionizing radiation
deposited.
The limitations of radiation damage can be greatly reduced by working
with specimens at cryogenic temperatures, as has been learned first in x-ray
crystallography \cite{low_pnas_1966,haas_acb_1970}, and later in
electron \cite{taylor_science_1974,dubochet_qrb_1988} and then x-ray
microscopy \cite{schneider_ultramic_1998,maser_jmic_2000}.  
This is done in part by reducing secondary chemical damage
effects caused by the radiolysis of water, as these radiolysis
products do not diffuse through ice at liquid nitrogen temperatures in
the same way as they do in water at room temperature.  An additional
benefit of cryogenic conditions is that scissioned molecular fragments
remain in place, unlike with room temperture specimens in vacuum
\cite{tinone_jvsta_1995} or in solution.
However, because cryogenic imaging conditions reduce but do not 
eliminate the limitations due to radiation damage, it remains
important to understand the relative radiation dose associated with
electron and x-ray microscopy for various specimen types.

\subsection{Comparisons of electron and x-ray microscopy}
\label{sec:past_comparisons}

The relative merits of electron and x-ray microscopy for transmission
imaging have been considered by several researchers who have reached
seemingly contradictory conclusions:
\begin{itemize}
\item \textbf{Molecular imaging can be done at significantly lower radiation
  dose in electron microscopy than in x-ray microscopy
  \cite{breedlove_science_1970,henderson_qrb_1995}.}  This conclusion
  is based on comparing fundamental scattering parameters for
  electrons as against those for x-rays while seeking to form images
  using elastically scattered quanta.  For 300 keV electrons incident
  on carbon, the inelastic
  scattering cross section is only $1.7\times$ larger than that for
  elastic scattering, and each 
  inelastic scattering event deposits only about 39 eV of energy as
  will be seen following Eq.~\ref{eqn:eels}; this means that only about
  66 eV of energy is deposited per elastic scattering event, or 1650
  eV if 25 elastically scattered photons are required to obtain a
  signal to noise ratio of 5.  This
  is about 500 times higher than the $\sim 3$ eV damage threshold for
  irreversible bond breaking in organic molecules, making it
  effectively impossible to obtain atomic resolution images of
  molecules
  \cite{breedlove_science_1970,glaeser_jur_1971,henderson_qrb_1995}
  (unless of course one combines data from low-dose images of many
  identical molecules \cite{frank_ultramic_1978,frank_science_1981,kuhlbrandt_science_2014}).
  However, $\sim 66$ eV per molecule per elastically scattered electron is dramatically
  lower than what happens with 10 keV x-rays, where the photoelectric
  absorption cross section is about 13 times higher than the elastic
  scattering cross section for carbon, and moreover photoelectric
  absorption involves the deposition of the entire energy of the x-ray
  photon; this means that about 130,000 eV of energy is deposited per
  elastically scattered x-ray photon.  The one exception to this
  pessimistic view of atomic resolution x-ray imaging might be if the entire
  x-ray scattering signal can be collected in tens of femtoseconds,
  during which time inertia might hold the molecule together
  \cite{solem_science_1982,neutze_nature_2000}.  While a similar
  femtosecond imaging approach has been contemplated using 
  electrons \cite{king_jap_2005}, Coulomb repulsion amongst
  charged particles represents a
  significant challenge \cite{glaeser_jsb_2008}.

\item \textbf{X-rays offer great advantages in penetration, so for
    studies of thicker biological materials it has been argued that
    soft x-ray microscopy can be carried out at a much lower radiation
    dose than is the case with electron microscopy
    \cite{sayre_ultramic_1977,sayre_science_1977}.}  These
  calculations showed that the greater penetrating power of x-rays was
  of great advantage for imaging hydrated organic specimens of the
  size of eukaryotic cells.  While they played an important role in
  motivating the development of x-ray microscopes, these earlier
  calculations left out several ingredients. In the case of x-ray
  microscopy, they overlooked the possibility of phase contrast, which
  greatly reduces contrast and reduces dose at multi-keV x-ray
  energies \cite{schmahl_xrmtaiwan} (this oversight was rectified in
  later calculations for x-ray microscopy
  \cite{golz_xrm1990,jacobsen_harder_xrm1990,jacobsen_xrm1996,schneider_ultramic_1998,wang_biotech_2013}).
  These early calculations
  \cite{sayre_ultramic_1977,sayre_science_1977} were done without
  awareness of the then-recent introduction of defocus phase contrast
  \cite{johnson_jrms_1968,unwin_jmic_1973} in electron microscopy.
  They also predated the introduction of zero-loss filtering, where an
  imaging energy filter is used to remove inelastically scattered
  electrons from the final image
 \cite{bauer_methmicrobio_1988,schroeder_jsb_1990} which otherwise
  contribute an out-of-focus image ``haze.''

\end{itemize}
In fact, these conclusions are neither contradictory, nor individually
incorrect as will be seen in Sec.~\ref{sec:new_comparison}.  It is our
purpose here to use the same calculation methodology for modern
approaches to electron and x-ray microscopy to re-evaluate the
relative advantages of these methods for different biological specimen
types, and also consider the ultimate thickness limits for x-ray
microscopy over a wide range of photon energies.

In the case of electron microscopy, several authors have considered
the improvements that phase contrast and zero-loss filtering can
provide for imaging thicker specimens.  In one case, Monte Carlo
numerical calculations were used to consider the fraction of electrons
that were scattered elastically and inelastically, or that had not
scattered at all, leading to estimates of an improved signal-to-noise
ratio for zero-loss imaging \cite{schroeder_jmic_1992}.  Increased
insight can be obtained by using not a numerical Monte Carlo model but
an analytical approach, since one can then find maxima of the
resulting mathematical functions.  One such approach added
consideration of electrons elastically scattered by large enough
angles that they are excluded by the objective lens aperture
\cite{langmore_ultramic_1992}.  Another approach \cite{grimm_bj_1998}
estimated the increased dose requirement in thicker specimens by
scaling a single projection image dose up by the number of tilts
required to maintain an equivalent 3D resolution in specimens of
increasing thickness \cite{hoppe_optik_1969} as quantified by the
Crowther criterion \cite{crowther_prsa_1970}.  However, for thicker
specimens one needs to consider plural elastic scattering within the
angle subtended by the objective lens aperture as another undesired
background signal, and we are unaware of calculations of image
contrast and required dose for electron microscopy that include this
effect beyond our own brief report \cite{jacobsen_xrm1996}.  In
addition, while this brief work included phase contrast in x-ray
microscopy, it did not consider the role of inelastic or plural
elastic x-ray scattering, effects that we include in this present
work.

\section{Calculation methodology}
\label{sec:calculation_methodology}

Our x-ray microscopy calculations are based on estimating image
intensities with features present and absent, following the method of
certain previous approaches for transmission imaging
\cite{glaeser_jur_1971,sayre_ultramic_1977,golz_xrm1990,jacobsen_harder_xrm1990,jacobsen_xrm1996,schneider_ultramic_1998,huang_optexp_2009}
as well as for x-ray fluorescence imaging
\cite{kirz_nyacad306,sun_ultramic_2015}. An alternative approach is to
consider the scattering strength of small features
\cite{solem_science_1982,solem_josab_1986,london_ao_1989,shen_jsr_2004,starodub_jsr_2008,howells_jesrp_2009,schropp_njp_2010}.
The results from these other calculations are broadly similar to the
thin specimen results from the method used here.  However, those other
calculations did not make the comparison with electron microscopy nor
did they include effects like plural elastic scattering or inelastic
scattering.  Estimations of the required exposure and dose have been
made for x-ray tomography at higher energies \cite{grodzins_nim_1983},
but those calculations predated the realization of the advantages of
phase contrast in x-ray tomography.

Our calculations here are for two dimensional images of features with a
thickness equal to the intended lateral resolution (that is, we assume
that small features have the same size in depth as they do laterally in a 2D
image).  However, they should apply equally well to 3D imaging on the
basis of using dose fractionation, which is the idea that ``a
three-dimensional reconstruction requires the same integral dose as a
conventional two-dimensional micrograph provided that the level of
significance and the resolution are identical''
\cite{hegerl_zn_1976}.  While originally controversial
\cite{hoppe_ultramic_1981}, this concept has entered practice and it
has withstood the test of
simulations \cite{mcewen_ultramic_1995}. In addition, it is an implicit
assumption in the successful method of single-particle imaging
\cite{frank_ultramic_1978,frank_science_1981,kuhlbrandt_science_2014}.

In what follows, we will start with more basic models and consider
their limits for the case of thin specimens, after which more complete
models will be discussed.  A guide to these various expressions is
provided in Table~\ref{tab:all_expressions}.

\begin{table}
  \begin{tabular}[c]{p{0.6\textwidth}c}
    \textbf{Contrast parameter} $\Theta$ 
       & \textbf{Equation number} \\
    X-rays, thin specimen, absorption contrast
      & Eq.~\ref{eqn:theta_abs_approx} \\
    X-rays, thin specimen, phase contrast, no phase ring absorption
       & Eq.~\ref{eqn:theta_zernike_approx} \\
    X-rays, thick specimen, absorption contrast, ignoring inelastic and plural
    elastic scattering  
      & Eq.~\ref{eqn:theta_abs} \\
    X-rays, thick specimen, phase contrast, ignoring inelastic and plural
    elastic scattering 
      & Eq.~\ref{eqn:theta_zernike} \\
    X-rays, unified complete expression 
       & Eq.~\ref{eqn:theta_unified} \\
%    X-rays, complete expression, dark field & Eq.~\ref{eqn:theta_df_complete} \\
    Electrons, phase contrast, without zero-loss energy filtering
       & Eq.~\ref{eqn:theta_nofilt} \\
    Electrons, phase contrast, with zero-loss energy filtering
       & Eq.~\ref{eqn:theta_filt} \\
    \end{tabular}
    \caption{Contrast parameters $\Theta$ for various imaging modes
      and degrees of approximation.  Once the contrast parameter
      $\Theta$ (defined in Eq.~\ref{eqn:theta}) and desired
      signal to noise ratio $\snr$ are specified, one can calculate
      the required exposure per pixel $\bar{n}$ using
      Eq.~\ref{eqn:min_n}, and the associated radiation dose using
      Eq.~\ref{eqn:x_dose} for x-rays or Eq.~\ref{eqn:d_e} for electrons.}
  \label{tab:all_expressions}
\end{table}

\subsection{Image statistics}
\label{sec:img_stats}

If all other noise sources are eliminated, the signal-to-noise ratio
of an image is dominated by photon statistics \cite{rose_jsmpe_1946}.
This is properly described by the Poisson distribution, which states
that an individual measurement will result in a probability distribution of
results $P(n,\bar{n})$ of
\begin{equation}
	P(n, \bar{n}) = \frac{\bar{n}^n}{n!}\exp(-\bar{n})
        \label{eqn:poisson}
\end{equation}
where $\bar{n}$ is the average over many measurements and $n$ is the
result of a particular measurement.  For values of $\bar{n}$ above
about 10, the Poisson distribution is well approximated by a Gaussian
distribution of 
\begin{equation}
	P(n, \bar{n}) =
        \frac{1}{\sqrt{2\pi\bar{n}}}\exp\Big[-\frac{(n-\bar{n})^2}{2\bar{n}}\Big]
\label{eqn:gaussian}
\end{equation}
which has a variance $\sigma^{2}$ characterized by
\begin{equation}
	\sigma = \sqrt{\bar{n}}.
	\label{eqn:sd}
\end{equation}
We then follow previous work
\cite{glaeser_jur_1971,sayre_ultramic_1977} and consider the signal to
be given by the unit-normalized image intensity $I_{f}$ at the
location of a feature $f$ if it is present, versus the intensity
$I_{b}$ of the background material $b$ at the same location if the
feature is absent.  If we illuminate this image pixel with $\bar{n}$
photons on average, the signal $\sig$ is then given by
\begin{equation}
	\sig = \bar{n}|I_f - I_b|.
        \label{eqn:signal}
\end{equation} 
The noise is the fluctuation due to photon statistics in both the
feature-present and feature-absent cases, since we are looking at the
difference between these two cases in Eq.~\ref{eqn:signal}.  The
statistical fluctuations of Eq.~\ref{eqn:sd} in these two measurements
are uncorrelated, so the noise $\noise$ is the root-mean-squared sum
of these two results or
\begin{equation}
	\noise = \sqrt{(\sqrt{\bar{n}I_f})^2 +
          (\sqrt{\bar{n}I_b})^2} = \sqrt{\bar{n}}\sqrt{I_f + I_b}.
        \label{eqn:noise}
\end{equation}
Consequently, the signal-to-noise ratio ($\snr$) is given by 
\begin{eqnarray}
  \snr &=& \sqrt{\bar{n}}\frac{|I_f-I_b|}{\sqrt{I_f+I_b}} \nonumber \\ 
  &=& \sqrt{\bar{n}}\Theta
 \label{eqn:snr}
\end{eqnarray}
where we define
\begin{equation}
  \Theta = \frac{|I_f - I_b|}{\sqrt{I_f + I_b}}
  \label{eqn:theta}
\end{equation}
as an image contrast parameter. Therefore
we obtain the result that the mean number of
quanta $\bar{n}$ with which we must illuminate each pixel in a 2D
image (or each voxel in a 3D tomographic reconstruction
\cite{hegerl_zn_1976}) is given by
\cite{glaeser_jur_1971,sayre_ultramic_1977}
\begin{equation}
  \bar{n} = \frac{\snr^{2}}{\Theta^{2}}.
  \label{eqn:min_n}
\end{equation}
Human vision
studies suggest a signal-to-noise ratio of $\snr=5$ leads to acceptable
image quality (the so-called ``Rose criterion''
\cite{rose_jsmpe_1946}), in which case we arrive at an expression of
\begin{equation}
  \bar{n} = \frac{25}{\Theta^2}
\end{equation}
for the minimum photon exposure per pixel.

\subsection{Specimen model}
\label{sec:specimen_model}

\begin{figure}[H]
%  \centerline{\includegraphics[width=0.4\textwidth]{pixel_slab}}
  \centerline{\includegraphics[width=0.4\textwidth]{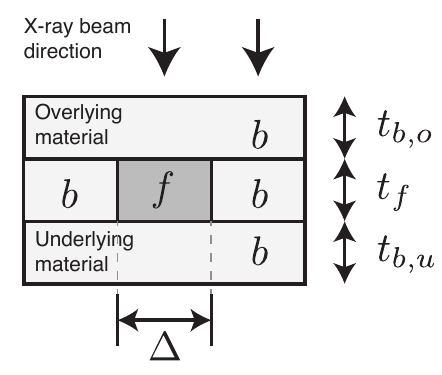}}
 \caption{Schematic diagram of the specimen model adopted for
    contrast parameter calculations. Features of material $f$ are
    assumed to be cubes of dimension $t_{f}$ on a side, embedded in an
    overall matrix thickness of $t=t_{b,o}+t_{f}+t_{b,u}$ of
    background material $b$.  For biological imaging, we will assume
    that the feature $f$ is composed of protein, and the background
    material $b$ is amorphous ice.}
  \label{fig:specimen_model}
\end{figure}

Our specimen model follows the assumption of several previous studies
\cite{glaeser_jur_1971,sayre_ultramic_1977}.
We consider the feature $f$ material to be in a cube of thickness $t_{f}$ which is
also equal to its transverse dimension (the spatial resolution in the
image).  It is contained within a matrix of a background material $b$
with thicknesses $t_{b,o}$ and $t_{b,u}$ over and under the feature,
respectively, so that the total specimen thickness $t$ is given by
\begin{equation}
  t = t_{b,o} + t_{f}+t_{b,u}
  \label{eqn:t_total}
\end{equation}
while the total for the underlying and overlying material is
\begin{equation}
  t_{b}=t_{b,o}+t_{b,u}
  \label{eqn:tb}
\end{equation}
as shown in Figure \ref{fig:specimen_model}.  Therefore $t_{f}$ might
represent the reconstructed voxel size in tomographic imaging, where
again we assume that dose fractionation applies \cite{hegerl_zn_1976}.

For biological materials we will follow the assumption that the
feature $f$ is comprised of the
stochiometric composition of a representative protein formed
from the average of all 20 amino acids, leading to a composition of
H$_{48.6}$C$_{32.9}$N$_{8.9}$O$_{8.9}$S$_{0.6}$ with a density when
dry of 1.35 g/cm$^{2}$ \cite{london_ao_1989}.  This likely slightly
overestimates the contrast of features $f$, since even densely-packed
macromolecules contain appreciable water. We will assume that the bulk
of the mass of a biological specimen is comprised of one of two
materials: amorphous ice with a density of 0.92 g/cm$^{3}$ for frozen
hydrated biological specimens \cite{dubochet_jmic_1982}, or the
embedding medium EPON for dehydrated specimens for which we assume a
stochiometric composition of \ce{C18H21O3Cl} and a density of 1.20
g/cm$^{3}$.

\section{Image contrast in x-ray microscopy}
\label{sec:xray_microscopy}

We first consider the case of image contrast and dose x-ray
microscopy.  We begin with a simplified model in
Sec.~\ref{sec:xray_contrast_simple} which follows earlier calculation
methods \cite{sayre_ultramic_1977,rudolph_modern_1990}.  We then
consider the additional factors of inelastic and plural elastic
scattering and arrive at a more complete model in
Sec.~\ref{sec:xray_contrast_complete}.

\subsection{X-ray microscopy: simplified model}
\label{sec:xray_contrast_simple}

Materials illuminated by x-ray beams have an index of refraction $n$
that is slightly less than unity
\cite{einstein_vdpg_1918,compton_philmag_1923} which can be written as
\begin{equation}
  n=1-\delta-i\beta.
  \label{eqn:xray_refractive_index}
\end{equation}
An alternative expression is
\begin{equation}
  n=1-\alpha \lambda^{2}(f_{1}+if_{2})
  \label{eqn:n_alpha_f}
\end{equation}
with
\begin{equation}
  \alpha \equiv \frac{r_{e}}{2\pi} n_{a},
  \label{eqn:alpha}
\end{equation}
where $r_{e}$ is the classical radius of the electron, $n_{a}$ is the
number density of atoms, and $(f_{1}+if_{2})$ describes the
frequency-dependent complex number of oscillator modes per atom in a
form for which excellent tabulations exist
\cite{henke_adndt_1993,elam_rpc_2002,schoonjans_sab_2011}.

The simplest contrast mechanism for x-ray imaging is absorption, for
which a transmitted beam has an intensity
\begin{equation}
  I=I_{0}\exp(-\mu t)
  \label{eqn:lambert_beer_law}
\end{equation}
according to the Lambert-Beer law.  Here $\mu$ is a linear absorption
coefficient that can be found from
\begin{equation}
  \mu=\frac{4\pi}{\lambda} \beta = 4\pi \alpha \lambda f_{2}.
  \label{eqn:lac}
\end{equation}
For the specimen model shown in Fig.~\ref{fig:specimen_model}, the
image intensity with a feature present is given by
\begin{eqnarray}
  I_\absf &=& I_0\exp(-\mu_f t_f)\exp(-\mu_b t_{b,o})\exp(-\mu_b
            t_{b,u}) \\ \nonumber
          &=& I_0\exp(-\mu_f t_f)\exp(-\mu_b t_b)
  \label{eqn:x_abs_f}
\end{eqnarray}
while the feature absent (background) case is given by
\begin{eqnarray}
  I_\absb &=& I_0\exp(-\mu_b t_{b,o}) \exp(-\mu_b t_f) \exp(-\mu_b
            t_{b,u}) \\ \nonumber
		&=& I_0\exp(-\mu_b t).
  \label{eqn:x_abs_b}
\end{eqnarray}
In the limits of $\mu_t t_f \ll 1$ and $\mu_{b}t_{f} \ll 1$, the above
equations can be simplified as
\begin{equation}
	I_\absf \approx I_{0} (1-\mu_f t_f)\exp(-\mu_b t_b)
	\label{eqn:ifabs}
\end{equation}
and
\begin{equation}
	I_\absb \approx I_0(1-\mu_b t_f)\exp(-\mu_b t_b).
	\label{eqn:ibabs}
\end{equation}
Letting $I_{0}=1$ for unit-normalized intensities in accordance with
the expression of Eq.~\ref{eqn:theta}, we arrive at a contrast
parameter $\Theta_{\textup{abs}}$ for x-ray absorption contrast
imaging of
\begin{eqnarray}
  \Theta_{\textup{abs}} &=& \frac{|I_\absf - I_\absb|}{\sqrt{I_\absf +
                          I_\absb}}\exp(-\mu_b t_b/2) \nonumber \\ 
    &=&
        \frac{|\exp(-\mu_{f}t_{f})-\exp(-\mu_{b}t_{f})|}{\sqrt{\exp(-\mu_{f}t_{f})+\exp(-\mu_{b}t_{f})}}
        \exp(-\mu_{b}t_{b}/2) \label{eqn:theta_abs}
\end{eqnarray}
which for thin specimens becomes
\begin{eqnarray}
   \Theta_{\textup{abs}} &\approx & \frac{t_f}{\sqrt{2}}|\mu_f - \mu_b|\exp(-\mu_b t_b/2) \label{eqn:theta_abs_approx} \\
     & \approx & \frac{2\pi \sqrt{2}}{\lambda}\, t_{f}\,
               |\beta_{f}-\beta_{b}|\exp(-\mu_{b}t_{b}/2) \label{eqn:theta_abs_approx2} 
\end{eqnarray}
as expressed using either the
linear absorption coefficients $\mu$ for
Eq.~\ref{eqn:theta_abs_approx} or the amplitude reduction part $\beta$
of the complex x-ray refractive index
(Eq.~\ref{eqn:xray_refractive_index}) for
Eq.~\ref{eqn:theta_abs_approx2}.

While it has long been known that the phase-shifting part $\delta$ of
the x-ray refractive index is much larger than the amplitude-reducing
part $\beta$, it was not until 1987 that Schmahl and Rudolph suggested
the use of phase contrast in x-ray imaging
\cite{schmahl_xrmtaiwan}. They considered the case of using the
Zernike method
\cite{zernike_physica_1934,zernike_physica_1942a,zernike_physica_1942b}
in x-ray microscopes, and arrived at expressions for the image
contrast \cite{rudolph_modern_1990} which we will briefly restate
here.  (There are a number of approaches to obtaining phase contrast
x-ray images which produce broadly similar results \cite{wilkins_ptrsa_2014}; the Zernike method has the
advantage of allowing for a direct calculation of image intensities as
required for the contrast parameter $\Theta$, and that is why we use
it here).  Zernike phase contrast in x-ray microscopy involves a linear phase coefficient
\begin{equation}
  \eta_{i} = 2\pi\frac{\delta_{i}}{\lambda}
  \label{eqn:eta}
\end{equation}
for phase advance per thickness in the materials of the feature
($i \rightarrow f$), the background ($i \rightarrow b$), and the phase
ring ($i\rightarrow p$). The original derivation began from
formulating the complex amplitudes $A$ of waves exiting from the
feature-present ($A_{f}$) and -absent ($A_{b}$) regions. We slightly
modify the equations to adapt them to the specimen model used in this
work, which contains an overlaying background layer on the dispersed
feature materials, leading to
\begin{eqnarray}
	A_f &=& A_0 \exp[-(\mu_b/2)t_b]\exp[i\eta_b t_b]\exp[-(\mu_f/2)t_f] \exp[i\eta_f t_f]
                \label{eqn:original_rudolph_af} \\
	A_b &=& A_0 \exp[-(\mu_b/2)t_b]\exp[i\eta_b t_b]\exp[-(\mu_b/2)t_f] \exp[i\eta_b t_f]. \label{eqn:original_rudolph_ab}
\end{eqnarray} 
The amplitude attenuation and phase shift
imposed upon the reference beam $A_{b}$ by transmission $T$ through the phase ring is
\begin{equation}
	T_p = \exp[-(\mu_p/2)t_p]\exp(i\eta_p t_p).
\end{equation}
As a result, the amplitude of the reference beam on the image plane is given by 
\begin{equation}
	A_b' = A_b T_p
\end{equation}
while the wave $A_{d}$ diffracted due to the presence of feature rather than
background material in the pixel $t_{f}$ is
\begin{equation}
	A_d = A_f - A_b.
\end{equation} 
Because of its scattering throughout the lens aperure, the diffracted
wave $A_{d}$ is nearly unaffected by the modulation $T_p$ of the small
phase ring.  As a result, the amplitude of the wave that
transmits through feature-present regions of the specimen takes the following
form at the image plane:
\begin{equation}
	A_f' = A_b' + A_d = A_b' + (A_f - A_b).
\end{equation}
Therefore the detected image intensities for the feature-present and
-absent cases are given by
\begin{eqnarray}
	I_\zpcf &=& A_f A_f^* \nonumber\\
	&=& e^{-\mu_b(t_b-t_f)}\biggl [(1 + e^{\mu_p t_p})e^{-\mu_b
                  t_f} + e^{-\mu_f t_f} + \label{eqn:rudolph_if} \\ 
	&& \hphantom{e^{-\mu_b(t_b-t_f)}\biggl [} 2e^{-\mu_f t_f/2 - \mu_b t_f/2}\cos(\eta_f t_f - \eta_b t_f - \eta_p t_p) - \nonumber \\
	&& \hphantom{e^{-\mu_b(t_b-t_f)}\biggl [} 2e^{-\mu_f t_f/2 - \mu_b t_f/2}\cos(\eta_f t_f -
          \eta_b t_f) - 2e^{-\mu_b t_f - \mu_p t_p/2}\cos(\eta_p
          t_p) \biggr]  \nonumber \\
	I_\zpcb &=& A_b A_b^* \nonumber\\
	&=& e^{-\mu_b(t_b-t_f)}e^{-\mu_b t_f - \mu_p t_p}. \label{eqn:rudolph_ib}
\end{eqnarray}
If the specimen is a weak phase object, and the phase ring produces a
phase shift of exactly $\pi/2$ with no absorption, these expressions
reduce to
\begin{eqnarray}
	I_\zpcf &\approx & I_0[1 + 2(\eta_f - \eta_b)t_f] \exp(-\mu_b
                           t_b) \label{eqn:ifzpc} \\ 
	I_\zpcb &\approx & I_0\exp(-\mu_b t_b). \label{eqn:ibzpc} 
\end{eqnarray}
We then have a contrast parameter for Zernike phase contrast imaging of
\begin{equation}
		\Theta_{\textup{zpc}} = \frac{|I_\zpcf -
                                          I_\zpcb|}{\sqrt{I_\zpcf +
                                          I_\zpcb}} \label{eqn:theta_zernike}
\end{equation}
involving the expressions of Eqs.~\ref{eqn:rudolph_if} and
\ref{eqn:rudolph_ib}.  In the weak phase contrast limits of $t_{f}\eta_{f} \ll 1$ and
$t_{f}\eta_{b} \ll 1$, this can be simplified by using
Eqs.~\ref{eqn:ifzpc} and \ref{eqn:ibzpc} to arrive at
\begin{eqnarray}
  \Theta_{\textup{zpc}} & \approx & \sqrt{2}\,t_f\, |\eta_f -
  \eta_b|\exp(-\mu_b t_b/2) \label{eqn:theta_zernike_approx} \\
  & \approx &  \frac{2\pi
    \sqrt{2}}{\lambda}\,t_{f}\,|\delta_{f}-\delta_{b}| \exp(-\mu_{b}t_{b}/2)
  \label{eqn:theta_zernike_approx2}
\end{eqnarray}
where Eq.~\ref{eqn:theta_zernike_approx2} emphasizes the symmetry with
Eq.~\ref{eqn:theta_abs_approx2}.

We have seen that the contrast parameters of the two imaging schemes
mentioned depend on the x-ray refractive indices
$1-\delta_{f}-i\beta_{f}$ for the feature material, and
$n=1-\delta_{b}-i\beta_{b}$ for the background material. With these
expressions in the thin specimen limit
(Eqs.~\ref{eqn:theta_abs_approx2} and
\ref{eqn:theta_zernike_approx2}), one can use Eq.~\ref{eqn:min_n} to
formulate simplified expressions for the number of photons $\bar{n}$
required for imaging at the Rose criterion of $\snr=5$ of
\begin{eqnarray}
	\bar{n}_\textup{abs} &=& \frac{25}{8\pi^{2}} \frac{\lambda^{2}}{t_f^{2}}\frac{1}{|\beta_{f} - \beta_{b}|^2}\exp(\mu_b t_b) \label{eqn:n_abs} \\ 
	\bar{n}_\textup{zpc} &=& \frac{25}{8\pi^{2}} \frac{\lambda^{2}}{t_f^{2}}\frac{1}{|\delta_{f} - \delta_{b}|^2}\exp(\mu_b t_b) . \label{eqn:n_zpc}
\end{eqnarray}
These expressions for the required number of photons per pixel scale
with feature thickness as $t_f^{2}$, while the area also decreases
with $t_f^{2}$ if we assume an isotropic (cubic) specimen; thus we are
consistent with other analyses
\cite{sayre_ultramic_1977,schneider_ultramic_1998,howells_jesrp_2009}
that arrive at a fourth power scaling between resolution improvements
and required exposure.

\section{X-ray microscopy: a more complete model}
\label{sec:xray_contrast_complete}

The above expressions are essentially what have been used in dose
estimates in x-ray microscopy that update the early work of Sayre
\emph{et al.}~\cite{sayre_ultramic_1977,sayre_science_1977} by
including phase contrast \cite{golz_xrm1990,jacobsen_harder_xrm1990}.
However, they are simplified expressions that ignore potentially
complicating effects beyond photoelectric absorption and simple
refractive phase.  We therefore wish to consider a more complete model.

\subsection{X-ray normalized intensity categories}
\label{sec:xray_normalized_intensities}

The probability $P$ for individual photon
interactions within a sample thickness $dt$ is given by
\begin{equation}
  P = \sigma_{i} \rho\, dt 
  \label{eqn:atomic_probability}
\end{equation}
where $\sigma_{i}$ is the cross section for interaction event
\textit{i}, and $\rho$ is the sample density. The interaction events
$i$ we now want to consider are photoelectric absorption as before ($i
\rightarrow \PI$), elastic or Rayleigh scattering ($i\rightarrow \el$),
and inelastic (Compton) scattering ($i \rightarrow \inel$).  These
cross sections are well tabulated
\cite{hubbell_jpcrd_1975,hubbell_jpcrd_1980,elam_rpc_2002} and are available in the
subroutine library \emph{xraylib} \cite{schoonjans_sab_2011} which we
have used for our calculations.  With these tabulated data in hand, we
can simplify the algebra that follows by writing the various
interaction coefficients per sample thickness $dt$ from
Eq.~\ref{eqn:atomic_probability} as
\begin{eqnarray}
K_{\el} &=& \sigma_{\el}\rho \label{eqn:kel} \\
K_{\inel} &=& \sigma_{\inel}\rho \label{eqn:kinel} \\
K_{\elin} &=& \sigma_{\el}(1-\eta_\el)\rho \label{eqn:kelin} \\
K_{\inelin} &=& \sigma_{\inel}(1-\eta_\inel)\rho
\label{eqn:inelin} \\
K_{\out} &=& \sigma_{\el}\eta_\el\rho + \sigma_{\inel}\eta_\inel\rho \label{eqn:kout}\\
K_{\PI} &=& \sigma_{\PI}\rho \label{eqn:kpl}
\end{eqnarray}
where $\eta_\el$ and $\eta_\inel$ are the probabilities that a photon
is scattered more than 90$^\circ$ (that is, backscattered and thus
lost to the imaging system) in an elastic and inelastic scattering
event, respectively. The two fractions can be found
\cite{sun_ultramic_2015} by integrating their corresponding
differential cross sections over the forward direction
($\theta \le 90\degree{}$) to obtain
\begin{eqnarray}
 \sigma_{\rm el} &=& \int_{0}^{\pi/2} [F(\theta/\lambda)]^{2} \frac{r_{e}^{2}}{2} (1+\cos^{2}\theta)\,d\theta \label{eqn:sel_from_thompson} \\
 \sigma_{\rm inel} &=& \int_{0}^{\pi/2} [S(\theta/\lambda)]^{2} \frac{r_{e}^{2}}{2}\left[1+k(1-\cos\theta)\right]^{-2} \left[1+\cos^{2}\theta+\frac{k^{2}(1-\cos\theta)^{2}}{1+k(1-\cos\theta)}\right]\,d\theta
  \label{eqn:sinel_from_kn}
\end{eqnarray}
where Eq.~\ref{eqn:sel_from_thompson} integrates the differential
Thomson cross section for unpolarized radiation and the atomic form
factor $F(\theta/\lambda,Z)$, while Eq.~\ref{eqn:sinel_from_kn}
integrates the Klein-Nishina cross section for the relative incident
energy $k=E/m_{e}c^{2}$ and uses the incoherent scattering form factor
$S(\theta/\lambda,Z)$.  One can obtain numerical values for both
$\sigma_{\rm el}$ and $\sigma_{\rm inel}$ by numerical integration of
the respective formulae as tabulated in the subroutine library
\emph{xraylib} \cite{schoonjans_sab_2011}.

\begin{figure}[H]
%  \centerline{\includegraphics[width=1\textwidth]{unimatrix_fig_x_ice}}
  \centerline{\includegraphics[width=1\textwidth]{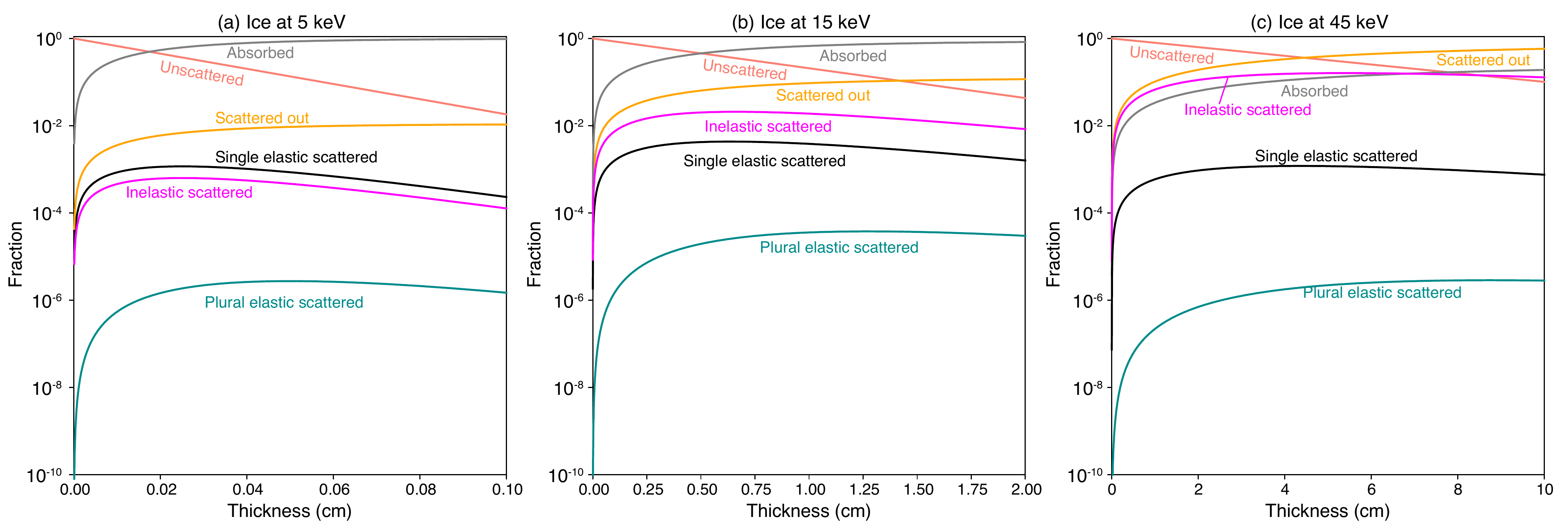}}
  \caption{Normalized intensity fractions for x-rays in amorphous ice as a
    function of thickness at incident photon energies of (a) 5 keV, (b) 15 keV,
    and (c) 45 keV. Phase contrast imaging involves an interference between 
    unscattered ($I_{\noscat}$; Eq.~\ref{eqn:x1}) and single elastically scattered
    ($I_{\sel}$; Eq.~\ref{eqn:x2}) photons, with other
    intensity fractions reprenting signal loss or
    background (these are described in Eqs.~\ref{eqn:x3}--\ref{eqn:x7}).  The corresponding intensity fractions for the
    embedding medium EPON are shown in Fig.~\ref{fig:epon_x_cate}.}
  \label{fig:ice_x_cate}
\end{figure}

\begin{figure}[H]
%  \centerline{\includegraphics[width=1\textwidth]{unimatrix_fig_x_epon}}
  \centerline{\includegraphics[width=1\textwidth]{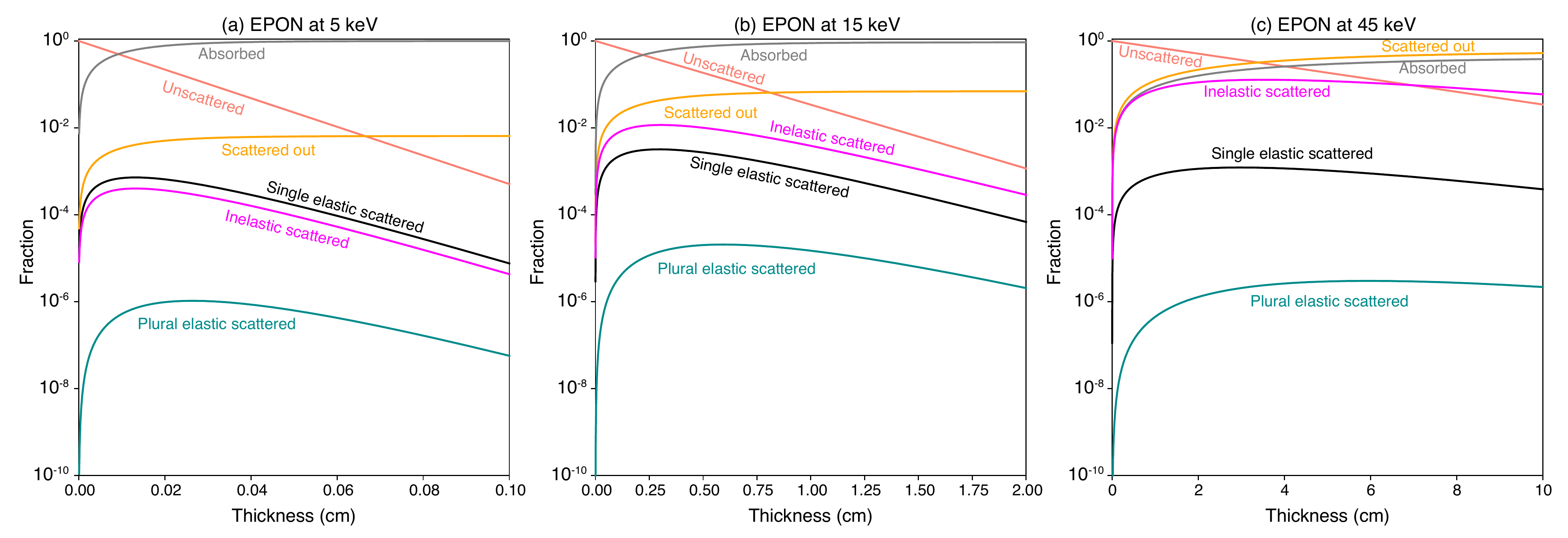}}
  \caption{Normalized intensity fractions for x-ray in EPON as a
    function of thickness at incident photon energies of (a) 5 keV, (b) 15 keV,
    and (c) 45 keV.  The corresponding intensity fractions for amorphous ice
    (such as for frozen hydrated specimes viewed under cryogenic
    conditions) are shown in Fig.~\ref{fig:ice_x_cate}. }
  \label{fig:epon_x_cate}
\end{figure}

With the above, we can now 
follow the example of calculations in electron
\cite{langmore_ultramic_1992} and x-ray \cite{jacobsen_xrm1996}
microscopy to consider a more complete set of categories to which
individual x-ray photons can ``join'' or ``leave'' depending on the
respective interaction cross sections:
\begin{itemize}

\item \textbf{Unscattered:} all photons are in this category
  at the outset, but they leave according to
\begin{equation}
dI_{\noscat} = -I_{\noscat}(K_{\inel}+K_{\el}+K_{\PI}) dt.
\label{eqn:x_di_noscat}
\end{equation}
The initial condition is $I_{\noscat}(0) = I_{\nt}$.

\item \textbf{Single elastic scattered:} unscattered photons can
  enter this category, while photons leave this category due to
  various scattering events giving
\begin{equation}
  dI_{\sel} = I_{\noscat}K_{\elin} dt -
  I_{\sel}(K_{\inel}+K_{\el}+K_{\PI}) dt
\label{eqn:x_di_sel}
\end{equation}
with $I_{\sel}(0) = 0$.

\item \textbf{Detected plural scattered:} photons undergoing multiple
  elastic scattering events while remaining within the detectable
  aperture are described by
\begin{equation}
  dI_{\elpl} = I_{\sel}K_{\elin} dt -
  I_{\elpl}(K_{\out}+K_{\inelin}+K_{\PI}) dt
\label{eqn:x_di_elpl}
\end{equation}
with $I_{\elpl}(0) = 0$.

\item \textbf{Backscattered:} photons that are scattered out
  of the detectable angular range (whether due to elastic, or
  inelastic, scattering) are given by
\begin{equation}
dI_{\out} = (I_{\nt} - I_{\out} - I_{\PI})K_{\out} dt
\label{eqn:x_di_out}
\end{equation}
with $I_{\out}(0) = 0$.

\item \textbf{Absorbed:} photons from any of the forward-directed
  \sout{category} \textcolor{blue}{categories} can always be absorbed according to
\begin{equation}
dI_{\PI} = (I_{\nt} - I_{\out} - I_{\PI})K_{\PI} dt
\label{eqn:x_di_Pl}
\end{equation}
with $I_{\PI}(0) = 0$.

\item \textbf{Not inelastic scattered:} the fraction of detectable
  photons that have not undergone inelastic scattering (the sum
  $I_{\noscat}+I_{\sel}+I_{\elpl}$) is denoted as $I_{\innoinel}$ and
  given by
\begin{equation}
dI_{\innoinel} = -I_{\innoinel}(K_{\inelin}+K_{\out}+K_{\PI})
\label{eqn:x_di_innoinel}
\end{equation}
with $I_{\innoinel}(0)=0$.

\item \textbf{Ineastically scattered within aperture:} the fraction of
  photons that undergo at least one inelastic scattering yet are still
  within the detectable aperture is given by
\begin{equation}
dI_{\inel} = I_{\innoinel}K_{\inel} dt - I_{\inel}(K_{\out}+K_{\PI})
dt
\label{eqn:x_di_inel}
\end{equation}
with $I_{\inel}(0) = 0$.

\end{itemize}
Solving these coupled differential equations yields
\begin{eqnarray}
  \mbox{Unscattered: } I_{\noscat} &=& I_{\nt}e^{-(K_{\inel}+K_{\el}+K_{\PI})t} \label{eqn:x1}
  \\ 
  \mbox{Single elastic scatttered: } I_{\sel} &=& I_{\nt}K_{\elin}e^{-(K_{\inel}+K_{\el}+K_{\PI})t} \\ \nonumber
	&=& K_{\elin}tI_{\noscat} \label{eqn:x2} \\
  \mbox{Plural scattered: } I_{\elpl} &=& I_{\nt} \Big[ e^{-(K_{\out}+K_{\inelin}+K_{\PI})t} - \nonumber \\
       &&\hphantom{I_\nt \Big[}(1+K_{\elin}t)e^{-(K_{\inel}+K_{\el}+K_{\PI})t} \Big] \label{eqn:x3} \\
  \mbox{Scattered out: } I_{\out} &=& \frac{I_{\nt}K_{\out}}{K_{\out}+K_{\PI}} \Big[1 -
             \exp[-(K_{\out}+K_{\PI})t]\Big] \label{eqn:x4} \\
  \mbox{Absorbed: } I_{\PI} &=& \frac{I_{\nt}K_{\PI}}{K_{\out}+K_{\PI}} \Big[1 -
             \exp[-(K_{\out}+K_{\PI})t]\Big] \label{eqn:x5} \\
  \mbox{Scattered in, no inelastic: } I_{\innoinel} &=& I_{\nt}\exp[-(K_{\inelin}+K_{\out}+K_{\PI})t] \label{eqn:x6} \\
  \mbox{Inelastic scattered: }I_{\inel} &=& I_{\nt}\Big[\exp[-(K_{\out}+K_{\PI})t] - \nonumber \\
              && \phantom{I_\nt\Big[}\exp[-(K_{\inelin}+K_{\out}+K_{\PI})t]\Big] \label{eqn:x7}
\end{eqnarray}
and it can be confirmed that the above expressions satisfy
\begin{equation}
I_{\noscat} + I_{\sel} + I_{\elpl} + I_{\out} + I_{\PI} + I_{\inel} = I_{\nt}
\end{equation}
as expected.  For a generic protein feature lying in amorphous ice and
EPON resin matrices, the expressions of
Eqs.~\ref{eqn:x1}--\ref{eqn:x7} are plotted in
Figs.~\ref{fig:ice_x_cate} and \ref{fig:epon_x_cate} as a function of
overall background material thickness for three example x-ray
energies: 5, 15, and 45 keV.

\subsection{Continuous versus atomistic features}
\label{sec:atomistic}

The above analyses can be used to evaluate the quality and statistical
quantities of the acquired images only if the continuous specimen
assumption remains valid.  This requires that neither the background
or feature material has optically significant structure at length
scales smaller than $t_{f}$, the spatial resolution of the imaging
experiment.  If that is the case, then there is a coherent
superposition of scattering amplitudes within the numerical aperture
of the imaging experiment from structures within the resolution scale
$t_{f}$.  This is in fact a condition assumed by the previous
imaging-based analyses cited earlier.  If instead there is significant
structure within a pixel, one might have to consider the fraction of
signal scattered to angles beyond the numerical aperture ($\NA$) of the
imaging system; that is, one would have a reduction in intensities
$I_\sel$, $I_\elpl$, and $I_\inel$. Obviously this approximation
becomes increasingly invalid as the resolution $t_{f}$ is decreased
down towards values where there are a small number of molecules within
a distance $t_{f}$ so that the feature begins to look ``lumpy'' and
scatters into larger angles. As an example, one simulation study of
gold atoms in amorphous and crystalline particles indicated that this
approximation becomes invalid at length scales of about 1 nm
\cite{dietze_jsr_2015}.

To understand the limits where atomic structure begins to produce
significant scattering beyond the acceptance of an experiment with a
limiting numerical aperture corresponding to the resolution $t_{f}$ in
biologically-significant materials, we turned to small angle x-ray
scattering (SAXS) data of 80 different protein samples retrieved from
Small Angle Scattering Biological Data Bank (SASBDB)
\cite{vaentini_nar_2015}. Data parameterization is done among all data
retrieved to yield a representative scattering distribution $I(s)$ of
\begin{equation}
	\log_{10}I(s) = 0.007466 s^5 - 0.1068 s^4 + 0.5305 s^3 - 0.8888 s^2 - 0.6426 s + 0.08601
	\label{eqn:saxs_poly}
\end{equation}
where
\begin{equation}
  s=4\pi\frac{\sin(\theta)}{\lambda}
  \label{eqn:momentum_transfer}
\end{equation}
is the momentum transfer of scattering (in nm$^{-1}$ for the
parameterization of Eq.~\ref{eqn:saxs_poly}), with $2\theta$ as the
scattering angle.  While it is usual to denote momentum transfer with
$q$ in the x-ray scattering literature, we use $s$ for consistency
with electron microscopy calculations \cite{langmore_ultramic_1992}.
From Eq. \ref{eqn:saxs_poly}, we can obtain an expectation value for
$s$ of
\begin{equation}
	\Braket{s} = \frac{\int_0^\infty s\,I(s)\,ds}{\int_0^\infty I(s)\,ds} \approx 0.42\ \mbox{nm}^{-1}.
\end{equation}
If we assume the value of the numerical aperture exactly matches
$\Braket{s}$, then the fraction $f_{\NA}$ of photons scattered into the
aperture can be found through
\begin{equation}
	f_{\NA} = \frac{\int_0^{\Braket{s}} I(s)\,ds}{\int_0^\infty I(s)\,ds} \approx 0.66.
\end{equation}
That is, approximately 66\%{} of the photons are accepted within a
square (rather than circular) numerical aperture if the imaging system
were to collect signal from a square of width
$\Delta=0.5 \lambda/\NA$ with $\NA$ replaced by $\Braket{s}=0.42$
nm$^{-1}$ (that is, a square pixel resolution of $\Delta=7.40$ nm
using 5 keV x-rays).  In other words, when the square pixel feature
size $t_{f}$ decreases towards a value of 7.40 nm, one would want
account for $f_{\NA}$ in the expressions for $I_{\sel}$,
$I_{\elpl}$, and $I_{\inel}$; however, we will not apply the
corrective factor $f_{\NA}$ in the calculations that follow.

\begin{figure}[H]
%  \centerline{\includegraphics[width=0.8\textwidth]{saxs_loglog}}
  \centerline{\includegraphics[width=0.8\textwidth]{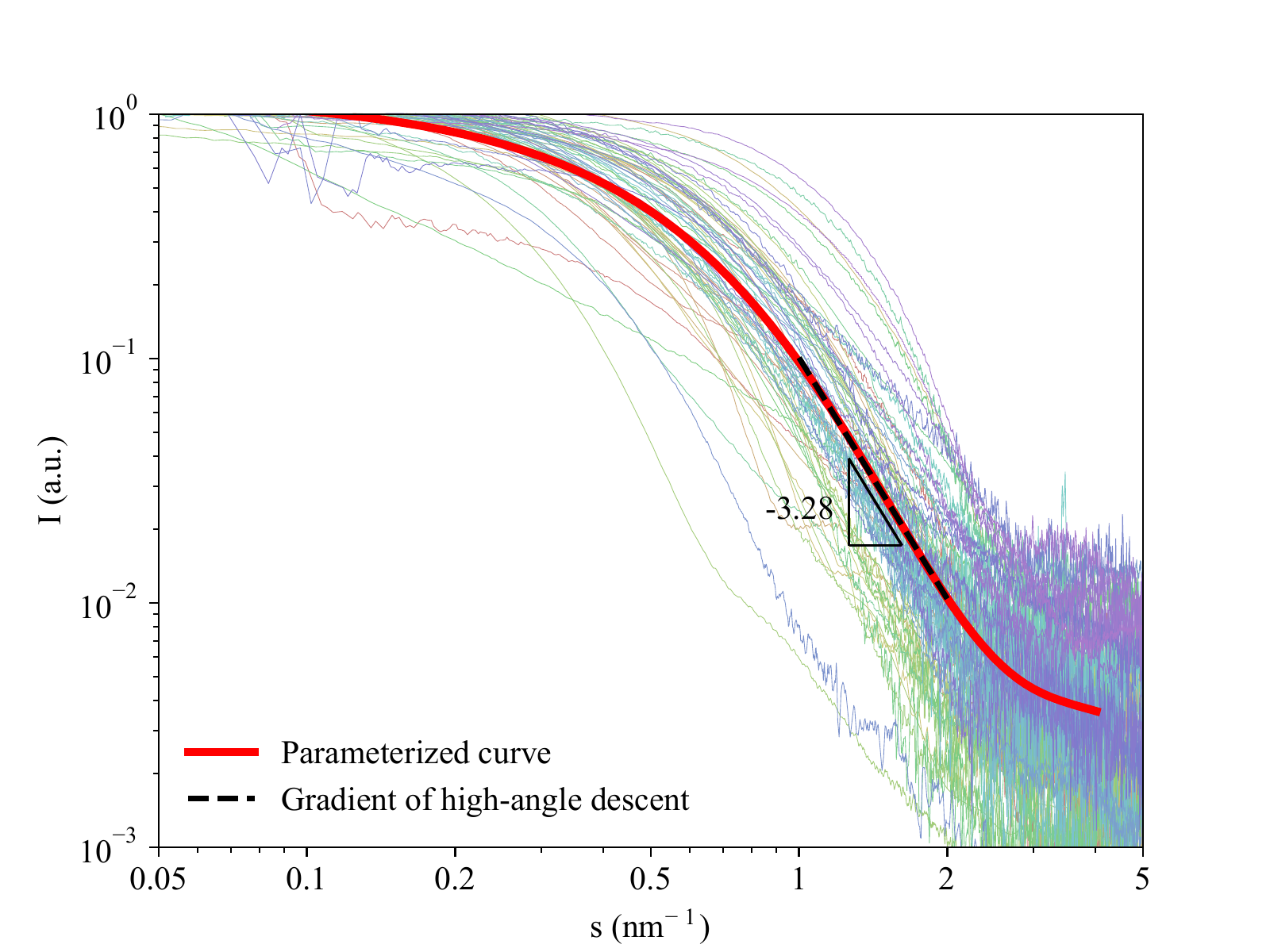}}
  \caption{The approximation of structure being continuous at length
    scales finer than the spatial resolution $t_{f}$ begins to break
    down when molecular scattering exceeds the corresponding numerical
    aperture.  Shown here as a series of thin lines are the small
    angle scattering patterns $I(s)$ from 80 different macromolecules
    in small angle scattering patterns in SASBDB, the Small Angle
    Scattering Biological Data Bank \cite{vaentini_nar_2015}.  To
    obtain a single parameterized representation of these patterns
    versus momentum transfer $s$, a sixth-order polynomial fit of
    $\log_{10}[I(s)]$ was obtained, leading to the expression of
    Eq.~\ref{eqn:saxs_poly}.  This parameterized fit is shown as the
    thick red line.}
  \label{fig:saxs}
\end{figure}

\subsection{X-ray imaging intensities for thicker specimens}
\label{sec:xray_imaging_intenstities}

Having determined the normalized intensities of x-rays in various
interaction categories, we can carry out a more complete calculation
of image signals with, and without, features present. Starting from
Eqs.~\ref{eqn:x1}--\ref{eqn:x7}, we arrive at expressions for the
background or feature-absent case of
\begin{eqnarray}
I_{\noscatb} &=& I_{\nt}\exp[-(K_{\inelb}+K_{\elb}+K_{\PIb})t] \label{eqn:xb1} \\
I_{\selb} &=& I_{\nt}K_{\elinb}\exp[-(K_{\inelb}+K_{\elb}+K_{\PIb})t] \nonumber \\ &=& K_{\elinb}t I_{\noscatb} \label{eqn:xb2} \\
I_{\innoinelb} &=& I_{\nt}\exp[-(K_{\inelinb}+K_{\outb}+K_{\PIb})t] \label{eqn:xb3} \\
I_{\elplb} &=& I_{\innoinelb} - I_{\noscatb} - I_{\selb} \label{eqn:xb4} \\
I_{\outb} &=& \frac{I_{\nt}K_{\outb}}{K_{\outb}+K_{\PIb}} \Big(1 - \exp[-(K_{\outb}+K_{\PIb})t]\Big) \label{eqn:xb5} \\
I_{\PIb} &=& \frac{I_{\nt}K_{\PIb}}{K_{\outb}+K_{\PIb}} \Big(1 - \exp[-(K_{\outb}+K_{\PIb})t]\Big) \label{eqn:xb6} \\
I_{\inelb} &=& I_{\nt} - I_{\outb} - I_{\PIb} - I_{\innoinelb} \label{eqn:xb7} %\\
%I_{\inelpcb} &=& I_{\inelb} \frac{\pi(\NA)^2}{2\pi} \label{eqn:xb8} \\
%I_{\elplpcb} &=& I_{\elplb} \frac{\pi\(NA)^2}{2\pi} \label{eqn:xb9}
\end{eqnarray} 
whereas the feature-present case is described by
\textcolor{blue}{equations including}
\begin{eqnarray}
  I_{\noscatf} &=& I_{\nt}\exp[-(K_{\inelb}+K_{\elb}+K_{\PIb})t_b] \exp[-(K_{\inelf}+K_{\elf}+K_{\PIf})t_f] \label{eqn:xf1} \\
  I_{\self} &=& (K_{\elinb}t_b + K_{\elinf}t_f) I_{\noscatf} \label{eqn:xf2} \\
  I_{\selff} &=& K_{\elinf}t_f I_{\noscatf} \label{eqn:xf2.5} \\
  I_{\innoinelf} &=& I_{\nt} \exp[-(K_{\outb}+K_{\inelinb}+K_{\PIb})t_b]
                     \exp[-(K_{\outf}+K_{\inelinf}+K_{\PIf})t_f] \label{eqn:xf3} \\
  I_{\elplf} &=& I_{\innoinelf} - I_{\noscatf} -
                 I_{\self}. \label{eqn:xf4}
\end{eqnarray}
\textcolor{blue}{The signal from the feature slice is gradually
  reduced in downstream layers due to scattering.  This is
  described by  $I_{\outff}$ and
  $I_{\PIff}$, which are formulated in a fashion similar to
  Eqs.~\ref{eqn:xb5} and \ref{eqn:xb6}. The differences within the
  feature slice are that the
  interaction coefficients (the
  $K$ coefficients) of the background material are replaced by those of the
  feature, and $t$ is changed to
  $t_f$. In addition, the incident intensity
  $I_\nt$ in Eqs.~\ref{eqn:xb5} and \ref{eqn:xb6} is replaced by
  $I_{\nt}-I_{\outb}(t_b/2)-I_{\PIb}(t_b/2)$, which is the intensity
  that remains after absorption and out-of-aperture scattering in the
  overlying slab (the ``incident intensity'' at the upper boundary
  of the feature slice):}
\begin{eqnarray}
  \textcolor{blue}{I_{\outff}} &=&
                 \textcolor{blue}{\frac{[I_{\nt}-I_{\outb}(t_b/2)-I_{\PIb}(t_b/2)]K_{\outf}}{K_{\outf}+K_{\PIf}} \Big(1 - \exp[-(K_{\outf} +K_{\PIf})t_f]\Big)} \label{eqn:xf5.0}  \\
  \textcolor{blue}{I_{\PIff}} &=&
                \textcolor{blue}{\frac{[I_{\nt}-I_{\outb}(t_b/2)-I_{\PIb}(t_b/2)]K_{\PIf}}{K_{\outf}+K_{\PIf}} \Big(1 - \exp[-(K_{\outf}+K_{\PIf})t_f] \Big).} \label{eqn:xf6.0}
\end{eqnarray}
\textcolor{blue}{Based on these, the total signal scattered outside the aperture of the
  detector is given by} 
\begin{align}
  \textcolor{blue}{I_{\outf}} &= \textcolor{blue}{I_{\outb}(t_b/2) + I_{\outff} +}   \tag{82a} \label{eqn:xf5} \\
  \nonumber & \qquad
                \textcolor{blue}{\frac{I_{\nt}-I_{\outb}(t_b/2)-I_{\PIb}(t_b/2)
                - I_{\outff} -
                I_{\PIff}}{I_\nt} I_{\outb}(t_b/2)}
\end{align}
\textcolor{blue}{The equation consists of three terms, describing the
  amount of photons scattered out in the overlaying background
  material, out of the feature slice in the middle, and out of the
  underlying background material. Following the definition in
  Eq.~\ref{eqn:xb5},
  $I_{\outb}(t_b/2)$ gives the amount of photons scattered out in the
  background material of thickness
  $t_{b}/2$, given an incident intensity of
  $I_\nt$, which is the thickness of the overlying slab. Within the
  feature slice at the middle, the amount of photons scattered out is
  $I_{\outff}$. The third term is the amount of out-of-aperture
  photons contributed by the underlying slab, but the incident
  intensity $I_\nt$ in
  $I_{\outb}(t_b/2)$ has to be replaced by the beam intensity after
  being attenuated by the overlying and middle slab, which is
  accounted for by the prefactor of $I_{\outb}(t_b/2)$. For
  $t_f \ll t_b$, the first and third terms in the equation can be collectively
  replaced by $I_{\outb}(t_b)$, resulting in }
\begin{equation}
  \textcolor{blue}{I_{\outf} \simeq I_{\outb}(t_b) + I_{\outff}}. \tag{82b} \label{eqn:xf5} 
\end{equation}
\textcolor{blue}{This approximation is based on the assumption that
the attenuation caused in the middle slice does not significantly
alter the beam intensity at the upper boundary of the underlying
material. Along with similar considerations for photoelectric
absorption, we arrive at}
\begin{align}
 \textcolor{blue}{ I_{\PIf}} &= \textcolor{blue}{I_{\PIb}(t_b/2) + I_{\PIff} +} \tag{82c} \label{eqn:xf6} \\
  \nonumber &  \qquad
                \textcolor{blue}{\frac{I_{\nt}-I_{\outb}(t_b/2)-I_{\PIb}(t_b/2)
                - I_{\outff} -
                I_{\PIff}}{I_\nt} I_{\PIb}(t_b/2)}
\end{align}
\textcolor{blue}{where the approximation at $t_f \ll t_b$ similarly applies to Eq.~\ref{eqn:xf6}.
Finally, this set of equations are completed by the expression}
\begin{equation}
\textcolor{blue}{I_{\inel}} = \textcolor{blue}{I_{\nt} - I_{\outf} - I_{\PIf} - I_{\innoinelf}.}\tag{82d} \label{eqn:xf7} 
\end{equation}

These intensities are dominated by interactions in the
background material as shown in Figs.~\ref{fig:ice_x_cate} and
\ref{fig:epon_x_cate}, except for the crucial differences caused by
feature material being present or absent in the region of width and
thickness $t_{f}$.

%I_{\PIf} &=&
%             \textcolor{blue}{\frac{[I_{\nt}-I_{\outb}(t_b/2)-I_{\PIb}(t_b/2)]K_{\PIf}}{K_{\outf}+K_{\PIf}} \Big(1 - \exp[-(K_{\outf}+K_{\PIf})t_f] \Big) + I_{\PIb}(t_b)} \label{eqn:xf6} \\

%I_{\inelpcf} &=& I_{\inelf} \frac{\pi(\NA)^2}{2\pi} \label{eqn:xf8} \\
%I_{\elplpcf} &=& I_{\elplf} \frac{\pi(\NA)^2}{2\pi} \label{eqn:xf9} \\

%Because the contrast parameter $\Theta$ of Eq.~\ref{eqn:theta}
%involves knowing signal differences due to having a thin feature
%present (or not) in the numerator, and backgrounds in the denominator,
%the incorporation of additional backgrounds due to thicker background
%materials $t_{b}$ will affect the denominator term of
%$\sqrt{I_{f}+I_{b}}$ only.  We can correct for this by introducting a
%correction factor $C_{a}(t)$ onto each of the intensities $I_{f}$ and
%$I_{b}$ of
%\begin{equation}
%  C_{a}(t) = 1+ I_\PIi(t) + I_\elpli(t) + I_\ineli(t) -I_\out(t)
%  \label{eqn:cat}
%\end{equation}
%based on the total specimen thickness $t$.  This has the consequence
%of modifying Eq.~\ref{eqn:theta_abs} to become
%\begin{equation}
%  \Theta_{\textup{abs}} =
% \frac{|\exp(-\mu_{f}t_{f})-\exp(-\mu_{b}t_{f})|}{\sqrt{C_{a}(t_{b})\big(\exp(-\mu_{f}t_{f})
%     +\exp(-\mu_{b}t_{f})\big)}} \exp(-\mu_{b}t_{b}/2)
%  \label{eqn:theta_abs_thick}
%\end{equation}
%for the case of background materials thick enough to produce
%significant inelastic and plural elastic scattering.
%\textcolor{red}{Ming, is this correct?}

\setcounter{equation}{82}
Now that we are considering the case where $I_\elpl$, $I_\inel$, and
$I_\out$ have non-negligible contributions, they have to be
incorporated into $\Theta$.  We do so by returning to the
feature-containing $A_{f}$ (Eq.~\ref{eqn:original_rudolph_af}) and
feature-absent $A_{b}$ (Eq.~\ref{eqn:original_rudolph_ab}) amplitudes
in Zernike phase contrast. These
amplitudes included a reduction $I_{0}[1-I_\PIi(t)]$ of the wavefield
incident upon the imaged pixel, but now the amplitudes are further
reduced by the presence of these additional scattering terms. This
means that these amplitudes must be reduced further according to a
correction term $C_{i}(t)$ of
\begin{equation}
 C_{i}(t) = \frac{I_0 - I_\PIi(t) - I_\elpli(t) - I_\ineli(t) -
   I_\out(t)}{I_0 - I_\PIi(t)}
 \label{eqn:czt}
\end{equation}
where $t$ again represents the total specimen thickness.  The modified
wave amplitudes corresponding to Eqs.~\ref{eqn:original_rudolph_af}
and \ref{eqn:original_rudolph_ab} then become
\begin{eqnarray}
	A_f &=& A_0 \exp[-(\mu_b/2)t_b]\exp[i\eta_b t_b]\exp[-(\mu_f/2)t_f] \exp[i\eta_f t_f]
               \sqrt{C_{b}(t_b) C_{f}(t_f)} \label{eqn:modified_rudolph_af} \\
	A_b &=& A_0 \exp[-(\mu_b/2)t_b]\exp[i\eta_b t_b]\exp[-(\mu_b/2)t_f] \exp[i\eta_b t_f] \sqrt{C_{b}(t_b) C_{b}(t_f)} \label{eqn:modified_rudolph_ab}
\end{eqnarray} 
where $A_0$ is the amplitude of the incident beam. If we neglect
absorption in the phase ring used in Zernike phase contrast so that it
only applies a phase shift $\exp[i\phi]$, the image intensity
equivalents to Eqs.~\ref{eqn:rudolph_if} and \ref{eqn:rudolph_ib} become
\begin{eqnarray}
	I_\signalf &=& 
	I_0 \exp(-\mu_b t_b)C_{b}(t_b) \biggl\{2\Bigl[\cos[\phi+(\eta_b-\eta_f)t_f] - \cos[(\eta_b - \eta_f)t_f]\Bigr] \times  \label{eqn:modified_rudolph_if} \\
	&& \hphantom{I_0 \exp(-\mu_b t_b)C_{b}(t_b) \biggl\{} \exp[-(\mu_b+\mu_f)t_f / 2] \sqrt{C_b(t_f) C_f(t_f)} + \nonumber \\
	&& \hphantom{I_0 \exp(-\mu_b t_b)C_{b}(t_b) \biggl\{} (2 - 2\cos\phi)\exp(-\mu_b t_f) C_b(t_f) + \exp(-\mu_f t_f) C_f(t_f)\biggr\} \nonumber \\ 
	I_\signalb &=& I_0 \exp(-\mu_b t)C_{b}(t_b) C_{b}(t_f) \label{eqn:modified_rudolph_ib} 
\end{eqnarray}
For an ideal phase ring with $\phi = \pi/2$, and the limiting case
that the amplitude correction terms $C_{i}(t)$ become 1,
Eqs.~\ref{eqn:modified_rudolph_if} and \ref{eqn:modified_rudolph_ib}
reduce to Eqs.~\ref{eqn:rudolph_if} and \ref{eqn:rudolph_ib}. If instead
we ``turn off'' the phase ring by setting $\phi = 0$ while also
working in the thin specimen limit with $C_{i}(t) \rightarrow 1$, it can be
shown that Eqs.~\ref{eqn:modified_rudolph_if} and
\ref{eqn:modified_rudolph_ib} reproduce the thin specimen limit
expresions for absorption contrast of Eqs.~\ref{eqn:x_abs_f} and
\ref{eqn:x_abs_b}.  Thus the intensities of
Eqs.~\ref{eqn:modified_rudolph_if} and \ref{eqn:modified_rudolph_ib}
can be used to describe both absorption ($\phi=0$) and Zernike phase
contrast ($\phi=\pi/2$) for thick specimens with the effects of
$I_\elpl$, $I_\inel$, and $I_\out$ included.  In this case the
contrast parameter $\Theta$ is modified from the expressions of
Eqs.~\ref{eqn:theta_abs} or \ref{eqn:theta_zernike} to become
\begin{equation}
 \Theta = \frac{|I_\signalf - I_\signalb|}{\sqrt{I_\pcf + I_\pcb}}.
 \label{eqn:theta_unified}
\end{equation}
using Eqs.~\ref{eqn:modified_rudolph_if} and
\ref{eqn:modified_rudolph_ib}, as well as an expression for the
background signals of
\begin{equation}
	I_\pci = I_\signali + I_\ineli + I_\elpli.
	\label{eqn:i_noise_unified}
\end{equation}
Because the plural elastic and inelastic terms do not carry structural
information, they only contribute to the noise term in the denominator
of Eq.~\ref{eqn:theta_unified}.

As summarized in Table~\ref{tab:all_expressions}, we have arrived at
three ways to express the contrast parameter $\Theta$ for absorption
and phase contrast in x-ray microscopy: in the simplest thin specimen
limit (Eqs.~\ref{eqn:theta_abs_approx2} and
\ref{eqn:theta_zernike_approx2}), in a form for thicker specimens as
has been described previously (Eqs.~\ref{eqn:theta_abs} and
\ref{eqn:theta_zernike}), and in a complete form that includes for the
first time the effects of inelastic and plural elastic scattering
(Eq.~\ref{eqn:theta_unified}).  When do these expressions differ
in significant ways?  Some insight can be provided by the relative
intensities shown in Figs.~\ref{fig:ice_x_cate} and
\ref{fig:epon_x_cate}, where it is clear that that $I_\noscat$ and
$I_\sel$ are orders of magnitude larger than the background signals
$I_{\inel}$ and $I_{\elpl}$ even if most of the incident beam has been
absorbed.  This means that even when the incident illumination is
greatly attenuated, the remaining imaging signal is still quite
``clean'' with relatively little signal contamination.  This is why X
rays are so successful at imaging very thick specimens, even up to
entire organisms.  

\begin{figure}[tb]
%  \centerline{\includegraphics[width=1\textwidth]{zpc_abs_tf0p02_tb40v_0p5kev}}
  \centerline{\includegraphics[width=1\textwidth]{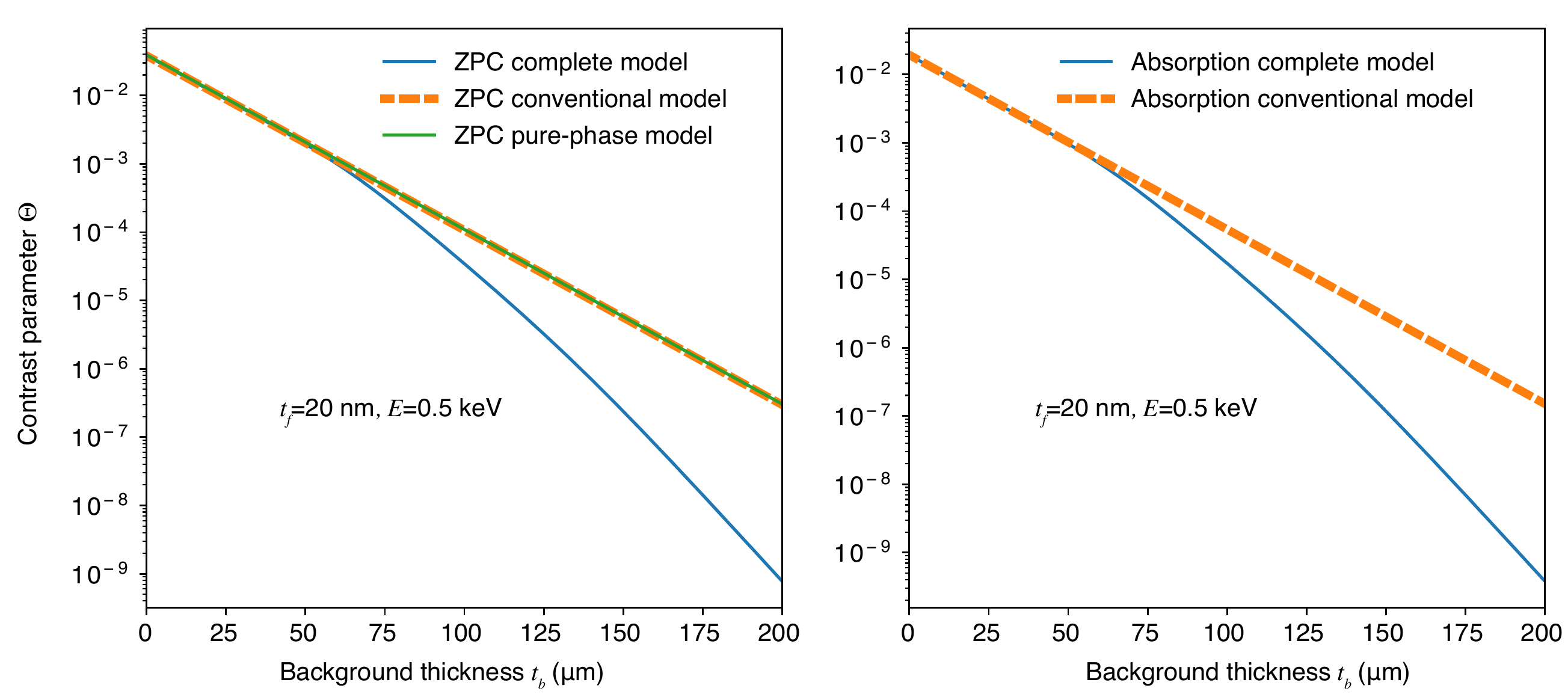}}
 \caption{\textcolor{blue}{Contrast parameter $\Theta$ for soft x-ray (0.5 keV)
    imaging of $t_{f}=20$ nm protein features as a function of
    amorphous ice thicknesses $t_{b}$.  At left is shown the
    case for Zernike phase contrast using the pure-phase thin
    sample approximation of Eq.~\ref{eqn:theta_zernike_approx},
    the conventional model of Eq.~\ref{eqn:theta_zernike}, and
    the complete model of Eq.~\ref{eqn:theta_unified} with phase
    contrast ($\phi=\pi/2$) .  The discrepancy between the
    pure-phase thin sample approximation and the conventional
    model is due to the fact that there is significant
    absorption at the soft x-ray energy of 0.5 keV, even though
    this is within the ``water window'' spectral region between
    the carbon (0.290 keV) and oxygen (0.540 keV) x-ray
    absorption edges.}}
  \label{fig:zpc_abs_tf0p02_tb40v_0p52kev}
\end{figure}

\begin{figure}[tb]
%  \centerline{\includegraphics[width=1\textwidth]{zpc_tf0p02_tf1000_tb1000v_15kev}}
  \centerline{\includegraphics[width=1\textwidth]{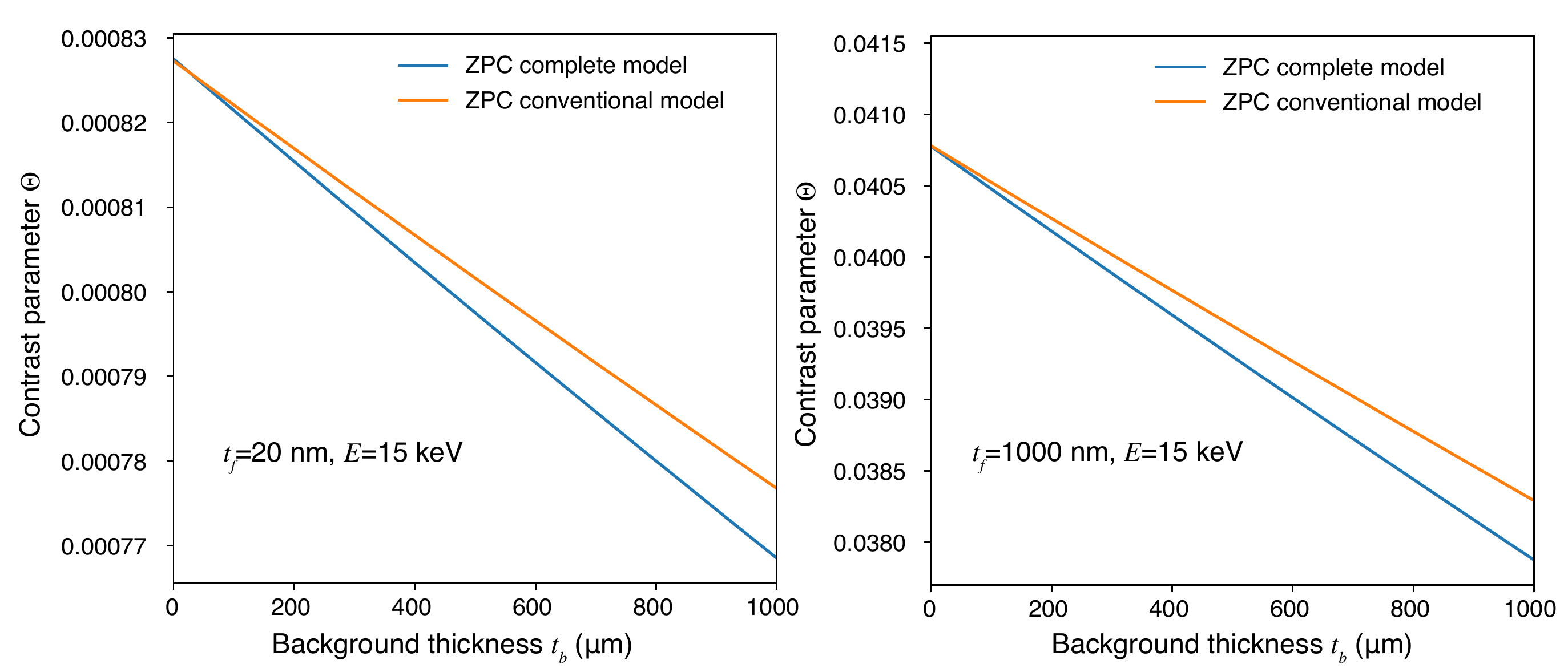}}
  \caption{\textcolor{blue}{Contrast parameter $\Theta$ for Zernike phase contrast
    imaging with hard x-rays (15 keV) as a function of overall
    amorphous ice thickness.  The case for a small protein feature
    ($t_{f}=20$ nm) is shown at left, while the case for a larger
    protein feature ($t_{f}=1000$ nm) is shown at right.  The
    conventional Zernike phase contrast model of
    Eq.~\ref{eqn:theta_zernike} works well for describing fine
    features in ice layers up to tens of micrometers thick, but the
    more complete model of Eq.~\ref{eqn:theta_unified} with phase
    contrast ($\phi=\pi/2)$ becomes necessary with thicker features
    and ice layers.  Absorption contrast is not shown because it is
    quite weak for hard x-ray imaging of organic materials in ice.}
	\label{fig:zpc_tf0.02_tf1000_tb1000v_15kev}}
\end{figure}

\begin{figure}[tb]
%  \centerline{\includegraphics[width=0.5\textwidth]{zpc_tf1000v_tb1000v_15kev}}
  \centerline{\includegraphics[width=0.5\textwidth]{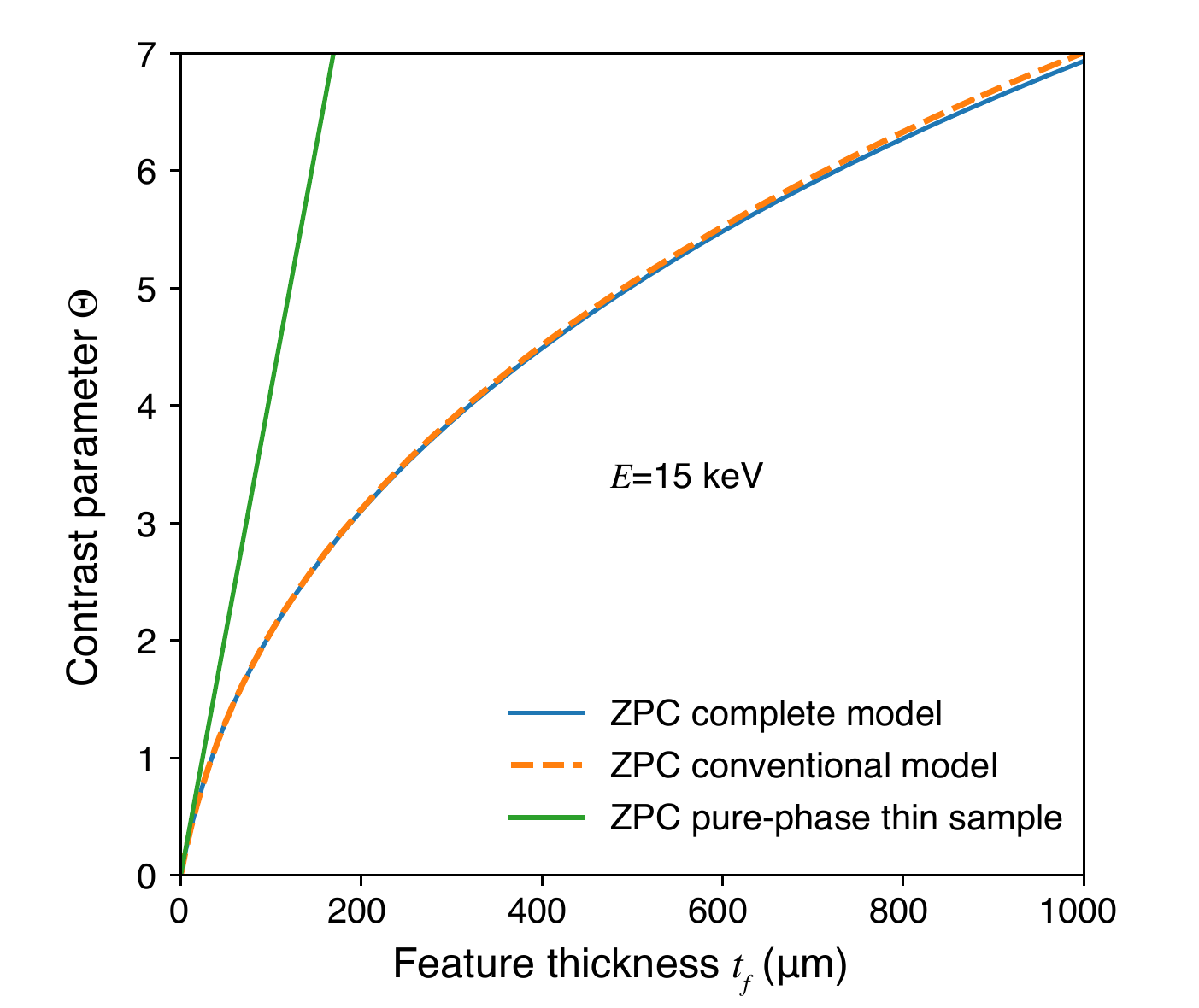}}
  \caption{\textcolor{blue}{Contrast parameter $\Theta$ for Zernike phase contrast
    imaging with hard x-rays (15 keV) as a function of feature
    thickness $t_{f}$ for protein in amorphous ice.  In this case no
    overlying or underlying thickness was assumed (that is,
    $t_{b,o}=t_{b,u}=0$ in Fig.~\ref{fig:specimen_model}), so this is
    just for a protein feature of the indicated thickness in an equal
    thickness slab of ice.  As can be seen, the pure-phase thin sample
    expression of Eq.~\ref{eqn:theta_zernike_approx} gives inaccurate
    predictions for ice thicknesses of even a few tens of micrometers;
    the conventional model of Eq.~\ref{eqn:theta_zernike} works well
    for thicknesses up to several hundreds of micrometers at which
    point the more complete expression of Eq.~\ref{eqn:theta_unified}
    with $\phi=\pi/2$ gives correct results.}}
	\label{fig:zpc_tf1000v_tb1000v_15kev}
\end{figure}

In order to see this more clearly, we show in
Figs.~\ref{fig:zpc_abs_tf0p02_tb40v_0p52kev}--\ref{fig:zpc_tf1000v_tb1000v_15kev}
the contrast parameter $\Theta$ calculated for different x-ray
energies, imaging conditions, and specimen and background thicknesses.
The case of Zernike phase and absorption contrast for soft x-ray
microscopy is shown in Fig.~\ref{fig:zpc_abs_tf0p02_tb40v_0p52kev},
illustrating the fact that the pure-phase approximation used in
Eqs.~\ref{eqn:theta_zernike_approx} and
\ref{eqn:theta_zernike_approx2} is inaccurate in the soft x-ray range
due to its exclusion of absorption contrast effects.  When using hard
x-rays (15 keV in this example), protein features in ice show almost
no absorption contrast so in
Fig.~\ref{fig:zpc_tf0.02_tf1000_tb1000v_15kev} we show Zernike phase
contrast only for fine features ($t_{f}=20$ nm) as well as for thicker
features ($t_{f}=1000$ nm) in ice thicknesses up to 1 mm.  As can be
seen, the conventional model of Eq.~\ref{eqn:theta_zernike} works well
for very thin features in thin ice layers, but as the overall specimen
thickness becomes larger than tens of micrometers and/or the feature
size approaches 1 \micron, one must use the complete expression of
Eq.~\ref{eqn:theta_unified} with $\phi=\pi/2$.  In other words,
inelastic and plural elastic scattering affect image contrast (and
therefore required exposure and dose) for micrometer-scale features as
well as for overall specimen thicknesses of tens of micrometers and above.

%When transmission x-ray microscopy is applied for examining biological and especially hydrated samples, dark field imaging is frequently adopted as a strategy to enhance contrast. This imaging mode involves the use of a downstream aperture with a central beam stop to block unscattered photons in the primary beam. The contribution of $I_\noscat$ is effectively removed in this case, thus 
%\begin{equation}
%	I_\signaldff = I_\selpcf.
%\end{equation}
%It follows that the contrast parameter $\Theta_\textup{df}$ is given by
%\begin{equation}
%  \Theta_\textup{df} = \frac{|I_\selpcf - I_\selpcb|}{\sqrt{I_\pcf + I_\pcb}}.
%  \label{eqn:theta_df_complete}
%\end{equation}

\section{X-ray imaging absorbed dose}
\label{sec:xray_dose}

Once the appropriate expression for the contrast parameter $\Theta$
has been evaluated, one can arrive at the required number of photons
per pixel $\bar{n}$ using Eq.~\ref{eqn:min_n}.  
This provide us with the basis of estimating the radiation dose, which
is the energy deposited per mass.  Accounting for attenuation by the
overlying background material shown in Fig.~\ref{fig:specimen_model},
the flux per area incident upon the feature is given by
$\bar{n} \exp(-\mu_{b} t_{b,o})$.  The energy deposited per length
$dE/dx$ can be found
from Eq.~\ref{eqn:lambert_beer_law} as
\begin{equation}
  \frac{dE}{dx} = \bar{n} \frac{hc}{\lambda} \frac{dI}{dx} =
  \bar{n} \frac{hc}{\lambda} \mu
\end{equation}
where $E=hc/\lambda$ is the photon energy based on Planck's constant
$h$ and the speed of light $c$.  Since this is incident on a
feature with area $\Delta^{2}$ and density $\rho_{f}$, this leads to a
dose $D_{f}$ absorbed in the feature of
\begin{equation}
  D_{f} = \bar{n} \frac{hc}{\lambda} \frac{\mu_{f}}{\rho_{f} \Delta^{2}}
  \exp(-\mu_{b} t_{b,o}).
  \label{eqn:x_dose_general}
\end{equation}
When the pixel width is equal to the
feature thickness (or $\Delta =t_{f}$), this becomes 
\begin{equation}
  D_{f} = \bar{n} \frac{hc}{\lambda} \frac{\mu_{f}}{\rho_{f} t_{f}^{2}}
  \exp(-\mu_{b} t_{b,o}).
  \label{eqn:x_dose}
\end{equation}
for the case of x-ray imaging of cubic features. Since the required
number of incident photons $\bar{n}$ is found from the contrast
parameter $\Theta$ according Eq.~\ref{eqn:min_n}, we arrive at a dose
to the feature of
\begin{equation}
  D_{f} = \frac{\snr^2}{\Theta^2} \frac{hc}{\lambda} \frac{\mu_{f}}{\rho_{f} t_{f}^{2}}
  \exp(-\mu_{b} t_{b,o})
  \label{eqn:general_dose}
\end{equation}
where setting $\snr$=5 corresponds to the Rose
criterion \cite{rose_jsmpe_1946}.
In the thin specimen limits of Eqs.~\ref{eqn:theta_abs_approx2} and
\ref{eqn:theta_zernike_approx2}, and with the use of the Rose criterion, one
can alternatively write the dose for absorption and phase contrast in
terms of the refractive index $n=1-\delta-i\beta$ as
\begin{eqnarray}
  D_{\rm abs} & \simeq & \snr^{2} \frac{hc}{2\pi \rho_{f} t_{f}^{4}} \frac{\beta_{f}}{|\beta_{f}-\beta_{b}|^{2}}\exp(\mu_{b}t_{b,u})
  		 \label{eqn:d_abs_approx} \\ 
  D_{\rm zpc} & \simeq & \snr^{2} \frac{hc}{2\pi \rho_{f} t_{f}^{4}} \frac{\beta_{f}}{|\delta_{f}-\delta_{b}|^{2}}\exp(\mu_{b}t_{b,u})
         \label{eqn:d_zpc_approx}	
\end{eqnarray}
which again shows that the required radiation dose $D$ 
increases with the fourth power of improvements in spatial resolution
$t_{f}$.  If one instead uses $\delta=\alpha \lambda^{2} f_{1}$ and
$\beta=\alpha\lambda^{2} f_{2}$ from Eq.~\ref{eqn:n_alpha_f}, the
phase contrast expression becomes 
\begin{equation}
  D_{f} \approx  \snr^{2} \frac{hc}{2\pi \alpha \rho_{f}}
  \frac{1}{t_{f}^{4}} \frac{1}{\lambda^{2}}
  \frac{f_{2,f}}{|f_{1,f}-f_{1,b}|^{2}}\exp(\mu_{b}t_{b,u}).
  \label{eqn:d_zpc_approx2}
\end{equation}
At x-ray energies in the keV range, $f_{1}$ tends to approach a
constant value of the atomic number $Z$ for the element in question
while $f_{2}$ scales like $\lambda^{2}$ \cite{henke_adndt_1993}.  Thus
the \emph{decrease} in $D_{f}$ due to $f_{2}$ nearly exactly cancels
out the \emph{increase} in $D_{f}$ due to the $1/\lambda^{2}$ term in
Eq.~\ref{eqn:d_zpc_approx2}.  As a result, once absorption in the
over- and underlying background material becomes negligible, the
required dose shows very little dependence on the x-ray energy used;
this will become apparent in
Figs.~\ref{fig:xray_dose_energy_thickness_contour_10nm} and
\ref{fig:xray_dose_energy_thickness_contour_100nm}.  The required
fluence $\bar{n}=\snr^{2}/\Theta^{2}$, however, will increase at
higher x-ray energies due to decreases in the contrast parameter
$\Theta$.

\section{Image contrast and dose in electron microscopy}
\label{sec:electron_contrast}

Transmission electron microscopy has a long and enormously successful
history in biological imaging, so it is useful to perform a similar
analysis of its characteristics.  In fact, several analytical analyses exist (as
noted Sec.~\ref{sec:introduction}), but some of them ignore defocus
phase contrast \cite{sayre_ultramic_1977} while others
\cite{langmore_ultramic_1992,schroeder_jmic_1992} do not carry their
estimates through to the point of considering defocus phase contrast
image intensities with feature present and absent as needed for the
contrast parameter $\Theta$ in Eq.~\ref{eqn:theta}.

\subsection{Electron cross sections}
\label{sec:electron_cross_sections}

While electrons undergo both atomic scattering and a
refractive-index-like phase advance due to the inner potential
\cite{bethe_adp_1928,wyrwich_znat_1958,reimer_1993}, there is not a
direct equivalent to the extensive tabulations available for the x-ray
refractive index in all materials \cite{henke_adndt_1993}.  However,
there are convenient parameterizations of the primary interaction
coefficients \cite{langmore_ultramic_1992} which we use below.  These
parameterizations include a fraction $\eta_{\el}$ of elastically
scattered electrons which do not pass through the objective aperture
of
\begin{equation}
  \eta_\el \approx 1 - \frac{s_0}{10}.
  \label{eqn:eta_el}
\end{equation}
The cross section for elastic scattering $\sigma_{\el}$ is well
approximated by
\begin{equation}
	\sigma_\el = \frac{1.4\times 10^{-6}Z^{3/2}}{\beta^2}\left(1 -
        \frac{0.26Z}{137\beta}\right) \mbox{ nm}^2,
\end{equation}
where $Z$ is the atomic number and $\beta$ is
the velocity relative to the speed
of light $c$ given by
\begin{equation}
  \beta^2 = 1 - \left(\frac{m_{e}c^{2}}{V_{0} + m_{e}c^{2}}\right)^{2}
\end{equation}
for an electron of mass $m_{e}$ accelerated over a voltage $V_{0}$.
The cross section for inelastic scattering $\sigma_{\inel}$ is
approxmated by
\begin{equation}
	\sigma_\inel = \frac{1.5\times10^{-6}Z^{1/2}}{\beta^2}\ln(2/\theta_c)\ \mbox{nm}^2
	\label{eqn:e_inel}
\end{equation}
where
\begin{equation}
  \theta_c = \frac{\Braket{\Delta E}}{\beta^2(V_{0} + m_{e}c^2)}
\end{equation}
with
$\Braket{\Delta E}$ representing the mean energy loss of electron in the
media. As noted by Langmore and Smith, 
Eq.~\ref{eqn:e_inel} gives erroneous results for hydrogen so instead
one uses
\begin{equation}
  \sigma_{\inel,\ Z=1} = 8.8\ \left(\frac{\beta_{80\ {\rm kV}}}{\beta}\right)^{2}\ \mbox{nm}^{2}
  \label{eqn:e_inel_h}
\end{equation}
which scales empirical observations made at $V_{0}=80$ kV.  The value
of $\eta_\inel$ (the fraction of inelastically scattered electrons
blocked by the objective aperture) is assumed to be $\eta_{\inel}=0$
as inelastic electron scattering typically involves very small angles
\cite{williams_1996}; we will therefore ignore $\eta_{\inel}$ in what
follows.

\subsection{Mean energy loss $\Braket{\Delta E}$}
\label{sec:mean_energy_loss}

\begin{figure}[H]
%  \centerline{\includegraphics[scale=.5]{ice_epon_eels}}
  \centerline{\includegraphics[width=0.9\textwidth]{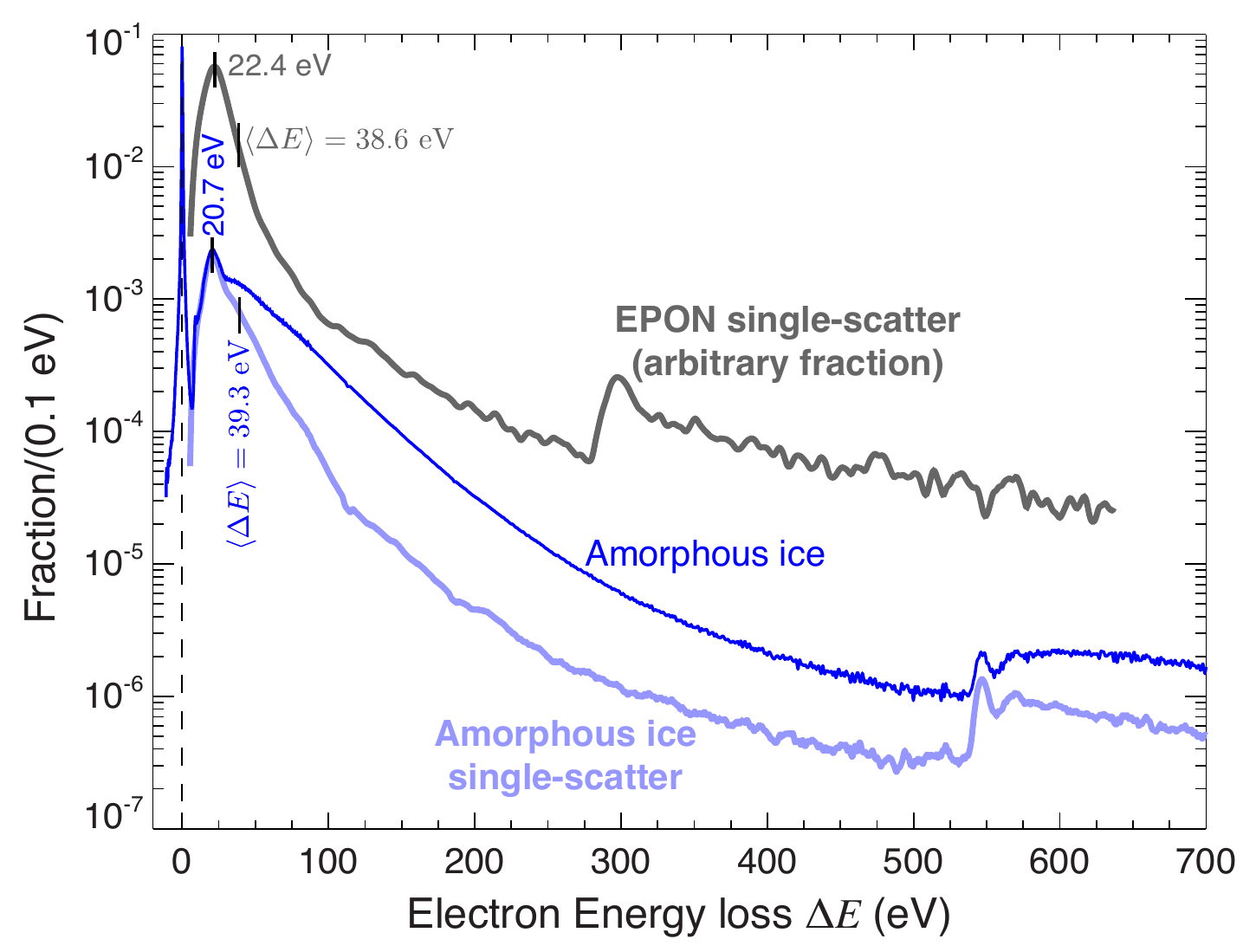}}
  \caption{Electron Energy Loss Spectroscopy (EELS) of amorphous ice
    and of the plastic embedding medium EPON. The EPON EELS spectrum
    shows a increase at the carbon $K$ edge at 290 eV, while the ice
    EELS spectrum shows an increase at the oxygen $K$ edge at 540
    eV. For amorphous ice, the as-recorded spectrum is shown along
    with the single-inelastic-scatter spectrum obtained by Fourier-log
    deconvolution; for EPON, only the single-inelastic-scatter
    spectrum is shown with arbitrary absolute scaling. Also shown are
    the locations of the plasmon mode peaks of the inelastic spectra,
    and the values of the mean energy loss $\Braket{\Delta E}$ as
    calculated using Eq.~\ref{eqn:eels}.  Amorophous ice spectra
    courtesy Richard Leapman, National Institutes for Health.  The
    EPON spectra are from a sample prepared by Qiaoling Jin, with
    assistance on EELS spectrum recording provided by Kai He, both of
    Northwestern University.}
  \label{fig:ice_epon_eels}
\end{figure}

The expression for the inelastic cross section $\sigma_{\inel}$
requires knowledge of the mean energy loss $\Braket{\Delta E}$ of
electrons in a material, and this quantity is also needed for
calculating the dose in electron microscopy (Eq.~\ref{eqn:d_e}).  This
energy can be measured using electron energy loss spectroscopy (EELS)
from a thin specimen.  We have obtained EELS spectra for our two
background materials: amorphous ice (based on 100 kV data provided by
Richard Leapman, National Institutes of Health), and EPON (based on 80
kV data acquired with Kai He, using a sample prepared by Qiaoling Jin;
both are with Northwestern University).  From the as-recorded
spectrum, one can calculate the single-inelastic-scatter spectrum
$I_{\inel}(\Delta E)$ using a Fourier-log deconvolution method
\cite{johnson_jpd_1974,wang_ultramic_2009}.  The raw EELS
spectrum for amorphous ice, as well as the single-inelastic spectra
for amorphous ice and for EPON, are shown shown in
Fig.~\ref{fig:ice_epon_eels}.  From the single-inelastic-scatter
spectrum $I_{\inel}(\Delta E)$, one can calculate the mean energy
loss $\Braket{\Delta E}$ using
\begin{equation}
	\Braket{\Delta E} = \frac{\int_0^\infty (\Delta E)
          I_\inel(\Delta E)\, d\Delta E}
           {\int_0^\infty I_{\inel}(\Delta E)\, d\Delta E}.
	\label{eqn:eels}
\end{equation}
The values we obtained for $\Braket{\Delta E}$ are very similar
between amorphous ice (39.3 eV) and EPON (38.6 eV), and these values
are similar to earlier measurements showing
$\Braket{\Delta E} \simeq 37$ eV in nucleic acids
\cite{isaacson_radres_1973,isaacson_siegel_1975}.

\subsection{Electron interaction probabilities}
\label{sec:electron_interactions}

\begin{figure}[H]
%  \centerline{\includegraphics[scale=0.5]{unimatrix_fig_e_ice}}
  \centerline{\includegraphics[width=0.9\textwidth]{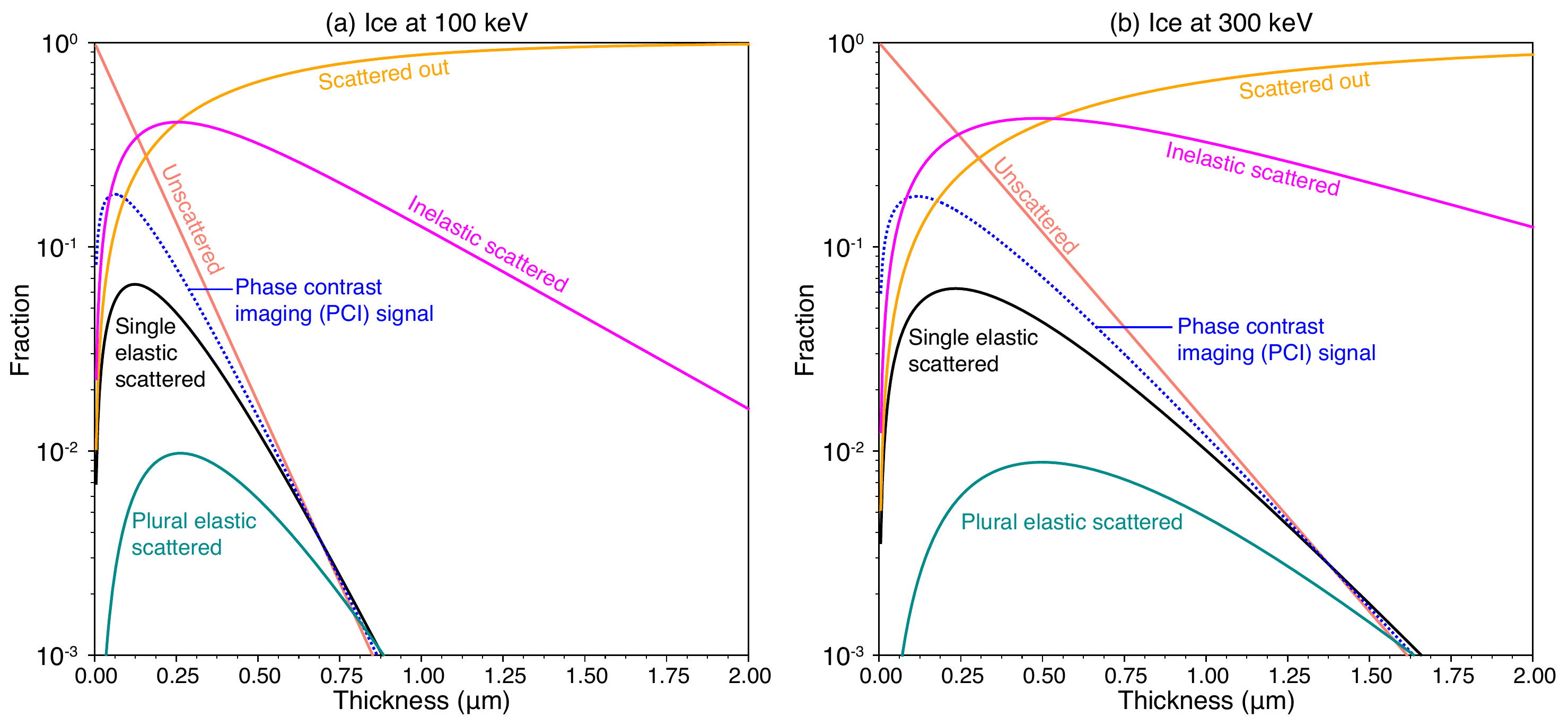}}
  \caption{Normalized intensity profiles for phase contrast electron imaging in amorphous ice
    as a function of thickness at incident electron energies of (a) 100
    and (b) 300 keV. The structural information in the image is contributed
    through the interference between unscattered electrons ($I_\noscat$, Eq. \ref{eqn:e1})
    and single elastic scattered electrons ($I_\sel$, Eq. \ref{eqn:e2}). 
    The intensities for EPON are shown in Fig. \ref{fig:epon_e_cate}.}
  \label{fig:ice_e_cate}
\end{figure}

\begin{figure}[H]
%  \centerline{\includegraphics[scale=0.5]{unimatrix_fig_e_epon}}
  \centerline{\includegraphics[width=0.9\textwidth]{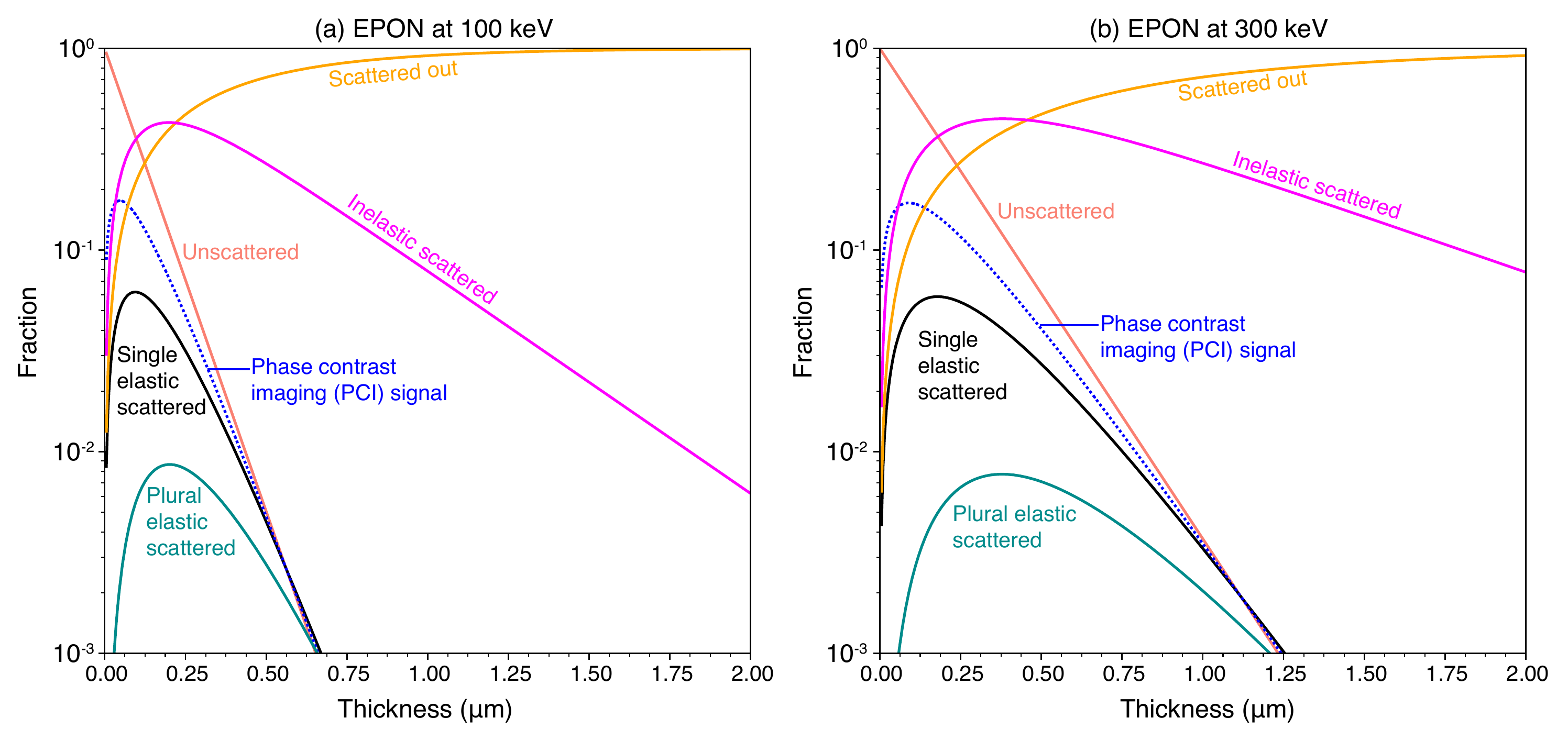}}
  \caption{Normalized intensity profiles for phase contrast electron imaging in EPON
    as a function of thickness at incident electron energies of (a) 100
    and (b) 300 keV. The intensities for amorphous ice are shown in Fig. \ref{fig:ice_e_cate}.}
  \label{fig:epon_e_cate}
\end{figure}

Having established functional forms for electron interaction cross
sections in Sec.~\ref{sec:electron_cross_sections}, we can proceed in
a manner similar to that used for x-ray interactions in
Sec.~\ref{sec:xray_contrast_complete}.  We first write the interaction
probability $K$ for electrons to undergo elastic scattering as
\begin{equation}
	K_\el = \sigma_\el \rho,
\end{equation}
for those singly scattered inside and outside the acceptance of the
objective respectively as
\begin{eqnarray}
  K_\elin &=& \sigma_\el (1 - \eta_\el) \rho \\
  K_\out &=& \sigma_\el \eta_\el \rho,
  \label{eqn:el_k_out}
\end{eqnarray}
and for the probability for electrons to be singly inelastically
scattered yet remain within the aperture as
\begin{equation}
   K_\inelin = \sigma_\inel \rho = K_\inel.
  \label{eqn:1inel_k_in}
\end{equation}
We can then assign electrons to categories \cite{jacobsen_xrm1996}
similar to those found in Eqs.~\ref{eqn:x1}--\ref{eqn:x7} to obtain
\begin{eqnarray}
  \mbox{Unscattered: } I_{\noscat} &=& I_{\nt}e^{-(K_{\inel}+K_{\el})} \label{eqn:e1} \\
  \mbox{Single elastic scattered: } I_{\sel} &=& I_{\nt}K_{\elin}e^{-(K_{\inel}+K_{\el})t} \nonumber \\ 
	 &=& K_{\elin}tI_{\noscat} \label{eqn:e2} \\
  \mbox{Plural scattered: } I_{\elpl} &=& I_{\nt} \Big[ e^{-(K_{\out}+K_{\inel})t} -
                (1+K_{\elin}t)e^{-(K_{\inel}+K_{\el})t} \Big] \label{eqn:e3} \\
  \mbox{Scattered out}: I_{\out} &=& I_{\nt} \Big(1 - e^{-K_{\out}t}\Big) \label{eqn:e4} \\
  \mbox{Scattered in, no inelastic}: I_{\innoinel} &=& I_{\nt}e^{-(K_{\inel}+K_{\out})t} \label{eqn:e6} \\
  \mbox{Inelastic scattered: } I_{\inel} &=& I_{\nt}\Big[e^{-K_{\out}t} -
              e^{-(K_{\inel}+K_{\out})t)}\Big]. \label{eqn:e7}
\end{eqnarray}
These normalized intensities for electrons in amorphous ice and in
EPON are shown in Figs.~\ref{fig:ice_e_cate} and
\ref{fig:epon_e_cate}, respectively, for the frequently-used electron
beam energies of 100 and 300 keV.  These plots can be compared with
the equivalents for x-rays shown in Figs. ~\ref{fig:ice_x_cate} and
\ref{fig:epon_x_cate}.  As noted in Sec.~\ref{sec:past_comparisons},
electron microscopy gives very strong interactions for thin materials
with modest energy deposition per elastically scattered electron, but
electrons have a much higher fraction of inelastic and plural elastic
scattered electrons present when imaging even micrometer thick
specimens; both of these signals can lead to a background ``fog''
when imaging thicker specimens.  With x-rays, the fractions of these
signals are much lower even for sample thicknesses of centimeters,
leading to considerably less image ``fog.''

\subsection{Electron image contrast}

The expressions of Eqs.~\ref{eqn:e1}--\ref{eqn:e7} were for a uniform
material.  We now consider our model specimen as shown in
Fig.~\ref{fig:specimen_model}, so that we can derive electron
microscopy equivalents to the x-ray microscopy expressions of Eqs.~\ref{eqn:ifabs}, \ref{eqn:ibabs},
\ref{eqn:ifzpc}, and \ref{eqn:ibzpc}.  When the pixel of interest
contains the background material, we obtain signal category
intensities of
\begin{eqnarray}
  I_{\noscatb} &=& I_{\nt}e^{-(K_{\inelb}+K_{\elb})t}. \label{eqn:eb1} \\
  I_{\selb} &=& I_{\nt}K_{\elinb}e^{-(K_{\inelb}+K_{\elb})t} \\ \nonumber
	&=& K_{\elinb}t I_{\noscatb} \label{eqn:eb2} \\
  I_{\innoinelb} &=& I_{\nt}e^{-(K_{\inelb}+K_{\outb})t} \label{eqn:eb3} \\
  I_{\elplb} &=& I_{\innoinelb} - I_{\noscatb} - I_{\selb} \label{eqn:eb4} \\
  I_{\inb} &=& I_{\nt}e^{-K_{\outb}t} \label{eqn:eb5} \\
I_{\inelb} &=& I_{\inb} - I_{\innoinelb} \label{eqn:eb7}
\end{eqnarray}
while when it contains the feature material we obtain
\begin{eqnarray}
I_{\noscatf} &=& I_{\nt}e^{-(K_{\inelb}+K_{\elb})t_b}e^{-(K_{\inelf}+K_{\elf})t_f} \label{eqn:ef1} \\
I_{\self} &=& (K_{\elinb}t_b + K_{\elinf}t_f) I_{\noscatf} \label{eqn:ef2} \\
I_{\selff} &=& K_{\elinf}t_f I_{\noscatf} \label{eqn:ef2.5} \\
I_{\innoinelf} &=& I_{\nt} e^{-(K_{\outb}+K_{\inelb})t_b} e^{-(K_{\outf}+K_{\inelf})t_f} \label{eqn:ef3} \\
I_{\elplf} &=& I_{\innoinelf} - I_{\noscatf} - I_{\self} \label{eqn:ef4} \\
I_{\inff} &=& I_{\nt}e^{-K_{\outb}t_b}e^{-K_{\outf}t_f} \label{eqn:ef5} \\
\textcolor{blue}{I_{\inelf}} &=& I_{\inff} - I_{\innoinelf}. \label{eqn:ef7} % Adding ",f" to the end
\end{eqnarray}
As noted in Sec.~\ref{sec:past_comparisons}, phase contrast dominates
in transmission electron microscopy of biological specimens.  Phase
contrast is usually obtained by using various defocus settings to
maximize contrast at different momenta transfer values $s$
\cite{johnson_jrms_1968,unwin_jmic_1973}, though there have been
advances in other approaches such as the Zernike method
\cite{glaeser_rsi_2013}.  An additional improvement in image contrast
\cite{bauer_methmicrobio_1988,schroeder_jsb_1990} can be obtained by
isolating the ``zero loss'' electrons using an image filter
(essentially an image-preserving electron monochromator located after
the objective lens); this excludes from the imaging plane the
inelastically scattered electrons which have a different kinetic
energy or de Broglie wavelength, so that they would otherwise
contribute out-of-focus information to the usual image plane.  We
therefore need to account for phase contrast and zero-loss imaging
using the intensity categories of Eqs.~\ref{eqn:eb1}--\ref{eqn:ef7}.

In electron microscopy, scattered electrons recieve a phase shift of
$\pi/2$ relative to the incident beam \cite{reimer_1993}; to maximize
the image intensity difference caused by the presence or absence of
these scattered electrons, their phase should be shifted by an
additional $\pm \pi/2$ before being recombined with the incident beam
at the location of the image pixel.  As a result, the image intensity
$I_{f}$ resulting from a feature being present in the specimen is
given by the interference between the unmodified incident wavefield
$\Psi_{\noscatf}$ and the $\pi/2$-phase-shifted wavefield from the
specimen feature $\Psi_{\self}$, leading to an image intensity of
\begin{eqnarray}
		I_{f} &= &|\Psi_\noscatf + \Psi_\self|^2 \\ \nonumber
		&= & I_\noscatf + I_\self + 2\sqrt{I_\noscatf I_\self}.
\end{eqnarray}
If instead there is no feature present, then the
background-material-containing pixel has no contrast variation within
itself or relative to the surrounding material
(Fig.~\ref{fig:specimen_model}), so its wavefield $\Psi_{\selb}$ will
undergo no additional phase shift due to defocus, Zernike, or other
phase contrast imaging methods in electron microscopy
\cite{johnson_jrms_1968,johnson_jmic_1973}.  As a result, the $\pi/2$
phase shift intrinsic to electron scattering will be unmodified, so
that the intensity recorded in the background-material image pixel
will be
\begin{eqnarray}
		I_{b} &=& \Big|\Psi_\noscatb + |\Psi_\selb| e^{i\pi/2}\Big|^2 \\ \nonumber
		&=& I_\noscatb + I_\selb.
\end{eqnarray}
We can therefore write the contrast parameter $\Theta$ of
Eq.~\ref{eqn:theta} as
\begin{equation}
%  \Theta_{\mathrm{e,unfiltered}} = \frac{|I_{\innoinelf}-I_{\innoinelb}|+2\sqrt{I_{\noscatf}I_{\selff}}}{\sqrt{I_{\inff}+I_{\inb}}}.
  \Theta_{\mathrm{e,unfiltered}} = \frac{|I_{\noscatf}-I_{\noscatb}+I_{\self}-I_{\selb}|+2\sqrt{I_{\noscatf}I_{\selff}}}{\sqrt{I_{\inff}+I_{\inb}}}.
  \label{eqn:theta_nofilt}
\end{equation}
when a zero-loss energy filter is absent, and
\begin{equation}
%  \Theta_{\mathrm{e,filtered}} = \frac{|I_{\innoinelf}-I_{\innoinelb}|+2\sqrt{I_{\noscatf}I_{\selff}}}{\sqrt{I_{\innoinelf}+I_{\innoinelb}}}.
   \Theta_{\mathrm{e,filtered}} = \frac{|I_{\noscatf}-I_{\noscatb}+I_{\self}-I_{\selb}|+2\sqrt{I_{\noscatf}I_{\selff}}}{\sqrt{I_{\innoinelf}+I_{\innoinelb}}}.
  \label{eqn:theta_filt}
\end{equation}
when a zero-loss energy filter is used. Note that the intensity
$I_\innoinel$ contributes to the background when zero-loss filtering
is not used, because these electrons will not be brought into a sharp
focus on the pixel in the imaging plane due to their different de
Broglie wavelength as noted above.  Instead, they will be distributed
in a more diffuse way on the detector plane, so that they effectively
make equal contributions to the intensities of both feature and
background pixels.

%\begin{align}
%\Theta_\nofilt &= \sqrt{I_\nt}\frac{(K_\elinf - K_\totalf + K_\totalb - K_\elinb)t_f + 2(1 - K_\totalf t_f)\sqrt{K\elinf t_f}}{\sqrt{2 - (K_\outf + K_\outb)t_f}} \label{eqn:theta_nofilt} \\
%\Theta_\filt &= \sqrt{I_\nt}\frac{(K_\elinf - K_\totalf + K_\totalb - K_\elinb)t_f + 2(1 - K_\totalf t_f)\sqrt{K\elinf t_f}}{\sqrt{2 - (K_\totalf - K_\elinf + K_\totalb - K_\elinb)t_f}} \label{eqn:theta_filt} 
%\end{align}

\subsection{Electron imaging absorbed dose}
\label{sec:electron_dose}

Once we have the required per-pixel illumination of
$\bar{n}=\snr^{2}/\Theta^{2}$ of Eq.~\ref{eqn:min_n} as calculated using either
Eq.~\ref{eqn:theta_nofilt} or Eq.~\ref{eqn:theta_filt}, we can
calculate the radiation dose imparted to the specimen based on the
fraction of inelastic scattering events and the mean energy
$\Braket{\Delta E}$ deposited per inelastic scatter
(Eq.~\ref{eqn:eels}).  The absorbed dose $D_\textup{e}$, or energy deposited per mass
in the pixel of area $\Delta^{2}=t_{f}^{2}$, involves the fraction of
interactions per length $K_\inelf$ in the feature of thickness
$t_{f}$ (Eq.~\ref{eqn:kinel}), leading to
\begin{equation}
	D_\textup{e} = \frac{\snr^{2}}{\Theta^2}\frac{\Braket{\Delta
            E}K_\inelf}{t_{f}^{2}\rho_f}.
	\label{eqn:d_e}
\end{equation}
Note that because the rejection of any large-angle-scattered electrons by
the objective aperture happens well downstream of the specimen plane,
we do \emph{not} include a term $[1-I_{\out}(t_{b,o})]$ which would
otherwise indicate a reduction in the beam intensity due to any
overlying background material.  Finally, we can use
Eqs.~\ref{eqn:kinel} and \ref{eqn:d_e}
to estimate a dose correponding to a given electron fluence; for
protein in amorphous ice, we find that a fluence of 1 $e^{-}$/nm$^{2}$
corresponds to a dose of $3.2\times 10^{4}$ Gray at 100 kV, and 
$1.8\times 10^{4}$ Gray at 300 kV.

\section{Implications for various imaging scenarios}
\label{sec:implications}

We have used a unified approach to estimate the required exposure and
radiation dose in x-ray and electron microscopy.  In this section, we
make use of this approach to consider both the comparison of x-ray and
electron microscopy, and to make several observations on the
characteristics of x-ray microscopy and imaging.  For these numerical
calculations, we will use the complete expression for x-ray microscopy
of Eq.~\ref{eqn:theta_unified}, and the expressions for electron
microscopy given in Eqs.~\ref{eqn:theta_filt} and \ref{eqn:theta_nofilt}
for the case of with and without zero-loss filtering, respectively.

\subsection{Comparison with experimental results}
\label{sec:experimental_comparison}

Before making these comparisons, we first wish to check our
calculations against some published experimental observations.  

While many electron microscopy and tomography papers quote the
electron fluence in $e^{-}$/nm$^{2}$, relatively few provide precise
quantitation of both the achieved spatial resolution and the overall
specimen thickness.  In addition, while in electron tomography one
should have nearly the same dose as a 2D image due to the dose
fractionation theorem noted above \cite{hegerl_zn_1976}, errors in the
alignment of projection images onto a common axis of rotation can
degrade image quality \cite{mcewen_ultramic_1995} and thus bias
reported results towards a higher electron fluence than might
otherwise have been necessary.  While there have been advances in
methods for tomogram alignment \cite{amat_methenz_2010}, details such
as the results of alignment convergence tests are almost never
reported.  In addition, single-particle imaging results are not
directly relevant to our comparison as they involve combining the
signals from independent images of a large number of identical
particles
\cite{frank_ultramic_1978,frank_science_1981,frank_arbbs_2002}.  For
these reasons, the best comparison is to consider 2D imaging of
thicker specimens.  In 120 kV zero-loss images of tobacco mosaic virus
in $t_{b}=100$ nm thick amorphous ice, fluence/resolution combinations
10 $e^{-}$/nm$^{2}$ at 4 nm, 300 $e^{-}$/nm$^{2}$ at 0.7 nm, and 1200
$e^{-}$/nm$^{2}$ at 0.4 nm resolution have been reported
\cite{grimm_jmic_1996}.  Our calculations provide estimates of
$\bar{n}=14$ at $t_{f}=4$ nm, $\bar{n}=470$ at $t_{f}=0.7$ nm, and
$\bar{n}=1430$ at 0.4 nm resolution, in all cases within 20--60\%{} of
the experimental observations.  We consider this to be a reasonable
agreement.

In x-ray microscopy, there are again only a limited number of examples
of 2D imaging of frozen hydrated specimens where the detection
efficiency is well known.  One such example involves 5.2 keV x-ray
ptychography as a method for phase contrast imaging of a frozen
hydrated cell with an overall ice thickness of $t_{b}=3$ \micron,
where a fluence of $9.2\times 10^{3}$ photons/nm$^{2}$ was used to
achieve sub-20 nm resolution \cite{deng_scirep_2017}.  Our calculation
provides an estimate of $\bar{n}=11.3\times 10^{3}$ photons/nm$^{2}$
at $t_{f}=20$ nm, again showing quite reasonable agreement.

\subsection{A comparison of x-ray and electron microscopy}
\label{sec:new_comparison}

In Section~\ref{sec:past_comparisons}, we noted that previous
comparisons of x-ray and electron microscopy arrived at seemingly
contradictory statements about their relative merits.  By using a
consistent methodology, we are able to address this question, building
upon previous work \cite{jacobsen_xrm1996} by including inelastic and
plural elastic scattering effects in x-ray microscopy as well as in
electron microscopy.

\begin{figure}[H]
%  \centerline{\includegraphics[width=0.9\textwidth]{dose_tf_10nm_bgthickness}}
  \centerline{\includegraphics[width=0.9\textwidth]{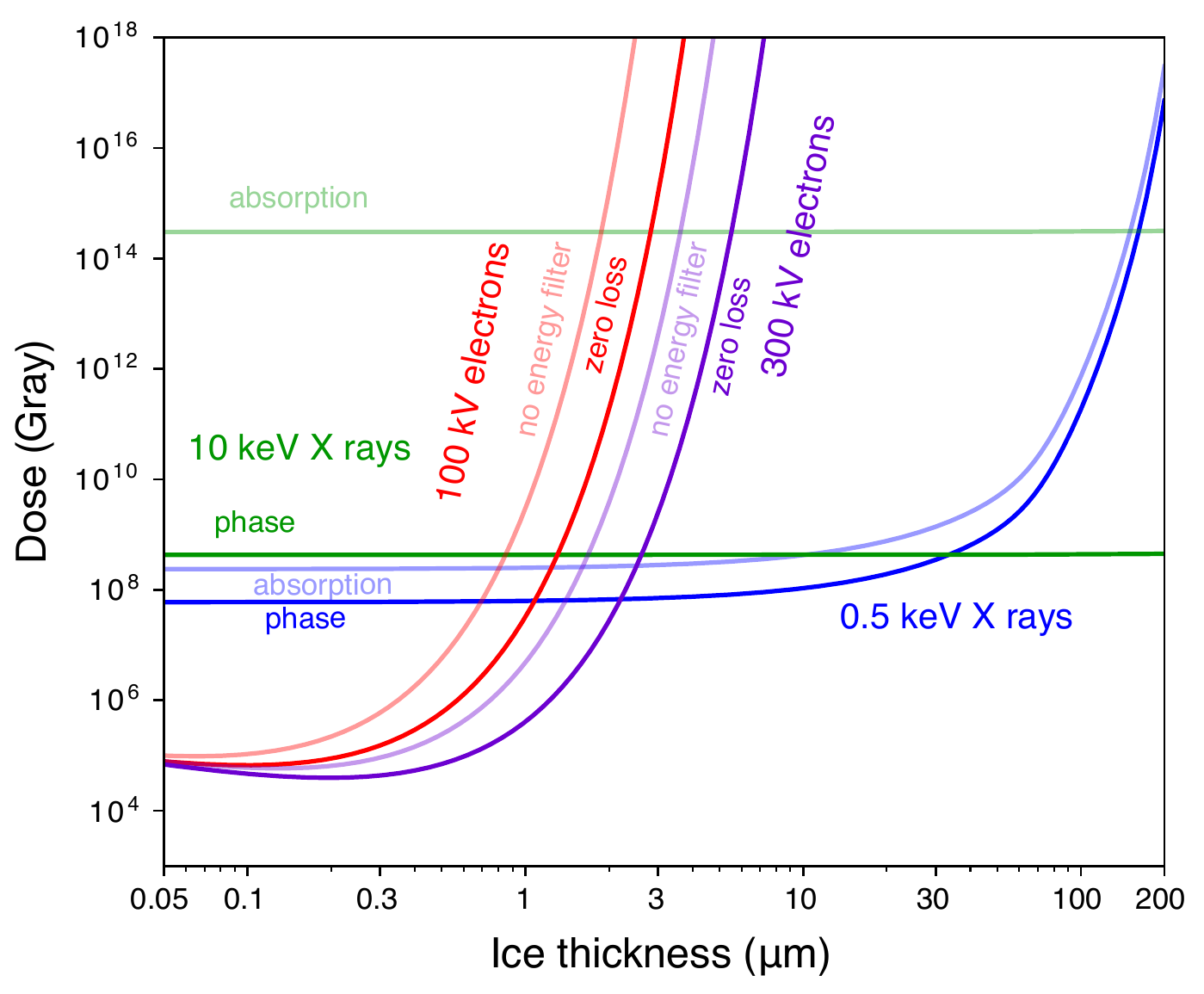}}
  \caption{\textcolor{blue}{Estimated radiation dose associated with 10 nm resolution
    imaging of protein features in amorphous ice.  This shows the case
    for soft x-ray microscopy at 0.5 keV and hard x-ray microscopy at
    10 keV, both for absorption and Zernike phase contrast.  In the
    case of electron microscopy, acclerating voltages of 100 and 300
    kV are shown for phase contrast imaging with and without the use
    of a zero-loss energy filter.  In all cases, the imaging system is
    assumed to have 100\%{} efficiency.}}
  \label{fig:electron_xray_dose_10nm}
\end{figure}

Because radiation dose scales strongly with spatial resolution, we
consider the case of a fixed spatial resolution of 10 nm which is
routine in electron microscopy and is within a factor of 2 of what is
now being demonstrated
x-ray microscopy of frozen hydrated biological specimens
\cite{deng_scirep_2017}.  (Even at slightly coarser resolution, x-ray
microscopy can be used to study structural changes in mitochondria associated
with virus infections \cite{perezberna_acsnano_2016}, while only
slight improvements in resolution are required for electron tomography
studies of actin networks
\cite{urban_ncb_2010}). Therefore we show in
Fig.~\ref{fig:electron_xray_dose_10nm} the estimated radiation dose
for imaging 10 nm protein features in amorphous ice thicknesses
ranging from 50 nm (typical of nicely-blotted plunge-frozen specimens
of macromolecules or virions), to $\sim 0.5$ \micron\ 
(typical of whole archaebacteria \cite{grimm_bj_1998} as well as
pheripheral regions of adherent eukaryotic cells
\cite{medalia_science_2002}), to thicknesses of 10 \micron\ 
(representative of the yeast \emph{S. cervisiae}
\cite{larabell_mbc_2004}), and beyond.  Examination of
Fig.~\ref{fig:electron_xray_dose_10nm} allows us to state the following:
\begin{itemize}

\item Electron microscopy offers dramatically lower radiation dose for
  high resolution imaging when the specimen thickness is much less
  than about 0.3 \micron, with extendability to slightly more than 1
  \micron\ for zero-loss imaging at 300 kV.

\item X-ray microscopy offers lower dose at thicknesses greater than
  about 1 \micron.  At the soft x-ray energy of 0.5 keV corresponding
  to the ``water window'' \cite{wolter_ap_1952} between the carbon $K$
  absorption edge at 0.29 keV and the oxygen $K$ edge at 0.54 eV,
  Zernike phase contrast provides some improvement over absorption
  contrast
  \cite{rudolph_modern_1990,golz_xrm1990,jacobsen_harder_xrm1990,schneider_ultramic_1998}
  in terms of minimum dose imaging.  At x-ray energies above the
  oxygen $K$ edge, absorption contrast becomes unfavorable; however,
  at energies greater than about 2 keV, phase
  contrast becomes quite favorable so that one
  can image features in amorphous ice layers as thick as several tens
  of micrometers \cite{schmahl_xrmtaiwan,golz_xrm1990,jacobsen_harder_xrm1990,wang_biotech_2013}.

\end{itemize}
In other words, Fig.~\ref{fig:electron_xray_dose_10nm} makes it clear
that the seemingly-contradictory prior conclusions outlined in
Sec.~\ref{sec:past_comparisons} are \emph{both} correct.

\begin{figure}[H]
%  \centerline{\includegraphics[width=0.9\textwidth]{xray_dose_energy_thickness_contour_10nm}}
  \centerline{\includegraphics[width=0.9\textwidth]{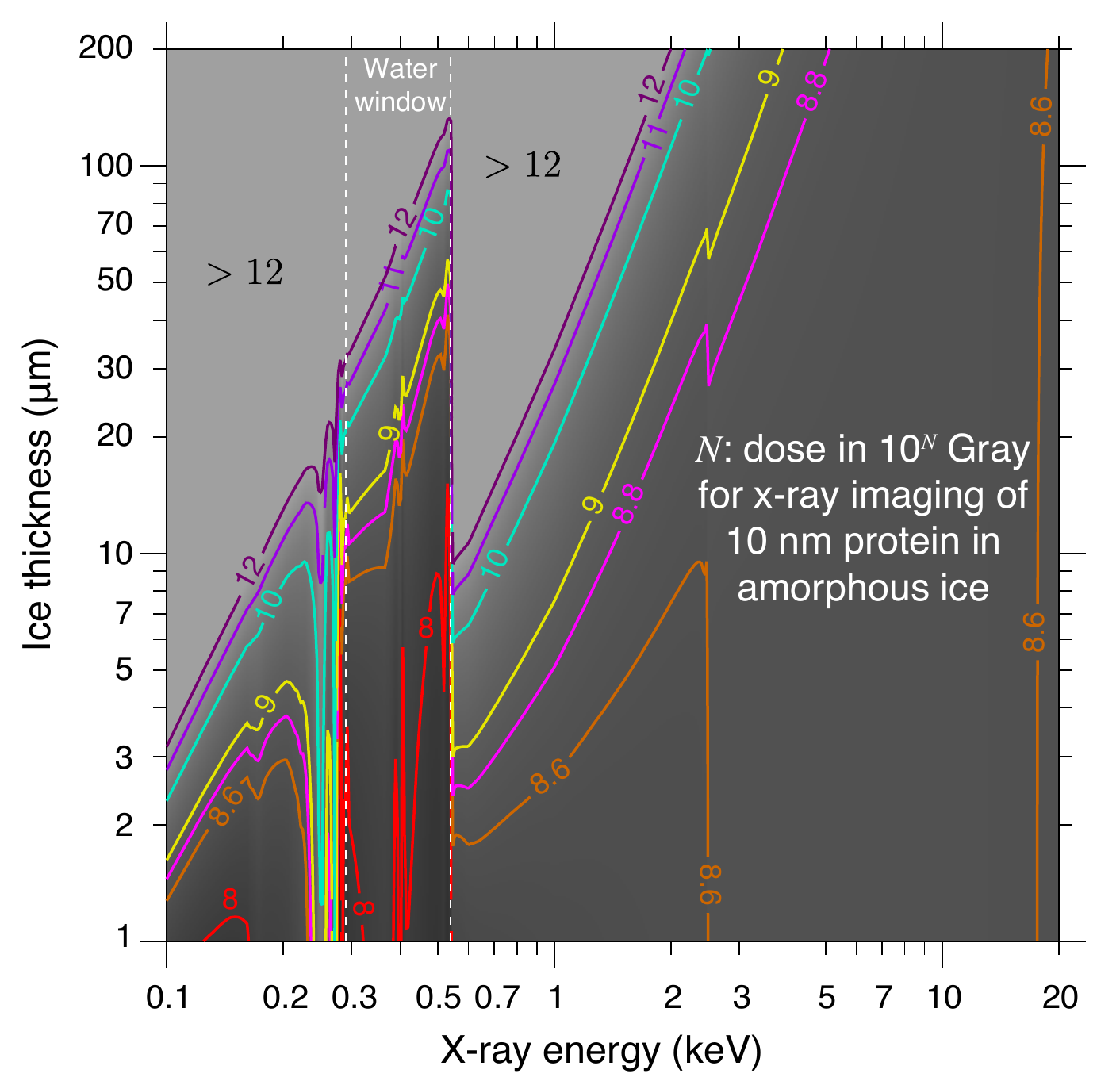}}
  \caption{\textcolor{blue}{Combined contour and brightness map of the required x-ray
    radiation dose in Gray for imaging 10 nm features in amorphous ice
    as a function of both x-ray photon energy and overall ice
    thickness.  This figure shows the lower of absorption or phase
    contrast imaging at each point; in nearly all cases phase contrast
    provides the lowest dose.  The grayscale image shows
    $\log_{10}({\rm Gray})$, with the overlaying contour line labeled
    6 representing a dose of $10^{6}$ Gray and so on.  The soft x-ray
    ``water window'' energy range \cite{wolter_ap_1952} between the
    carbon $K$ edge at 0.29 keV and the oxygen $K$ edge at 0.54 eV
    provides minimum dose imaging for specimens in ice layers up to
    about 10--20 \micron\ thick, while phase contrast requires a
    slightly higher dose at multi-keV energies while accommodating
    thicker specimens overall.  Note that the presence of sulfur in
    our model protein leads to the contour feature at the $S$ $K$ edge
    at 2.47 keV.}}
  \label{fig:xray_dose_energy_thickness_contour_10nm}
\end{figure}

In order to better understand the optimal photon energy to be used for
transmission imaging in x-ray microscopy, in
Fig.~\ref{fig:xray_dose_energy_thickness_contour_10nm} we show the
dose required for $\snr$=5 imaging of 10 nm protein features as a
function of both x-ray energy and overall amorphous ice thickness.
Once again, this figure shows the advantages of the 
``water window'' spectral region \cite{wolter_ap_1952}, as well as the
utility of phase contrast imaging at energies above about 2 keV.

\subsection{Dose versus resolution in x-ray microscopy}
\label{sec:dose_versus_resolution}

\begin{figure}[H]
%  \centerline{\includegraphics[width=0.8\textwidth]{dose_tb_10um_resolution}}
  \centerline{\includegraphics[width=0.8\textwidth]{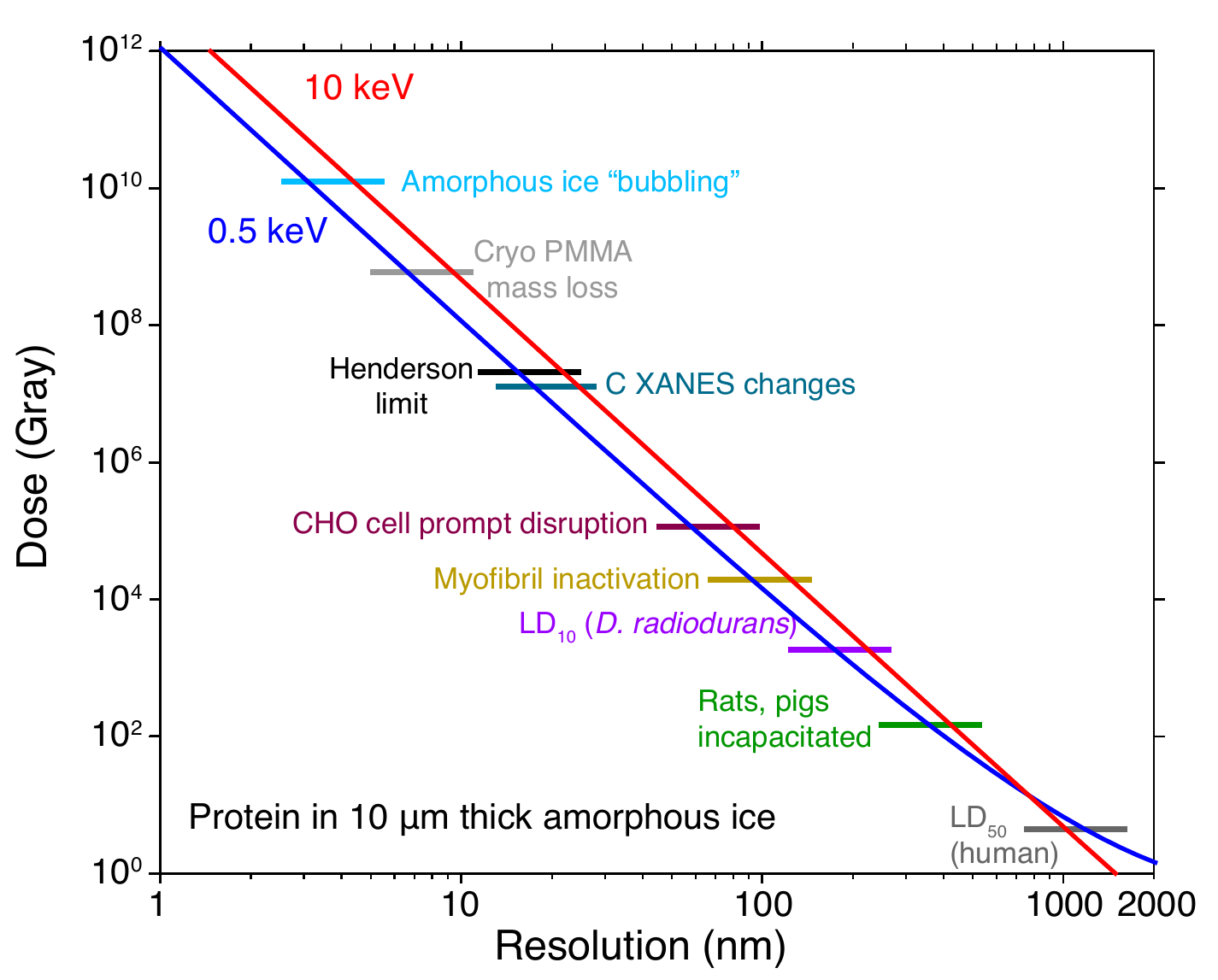}}
  \caption{\textcolor{blue}{Required radiation dose as a function of resolution.  In
    this case of imaging protein in 10 \micron\ of amorophous ice, the
    dose for $\snr$=5 imaging was calculated as $t_{F}$ was varied,
    for both soft x-rays in the water window (0.5 keV) or for hard X
    rays (10 keV), for the better of absorption or Zernike phase
    contrast at each thickness.  The trend of required dose increasing
    as the fourth power of improvements in spatial resolution
    (decreases in $t_{f}$) as expected from
    Eqs.~\ref{eqn:d_abs_approx} and \ref{eqn:d_zpc_approx} is clearly
    seen.  Also shown are the radiation doses associated with various
    detrimental effects in biological specimens, as discussed in
    Sec.~\ref{sec:dose_versus_resolution}.}}
  \label{fig:xray_dose_versus_resolution}.
\end{figure}

As discussed in Sec.~\ref{sec:xray_contrast_simple} and shown in the
approximate results of Eqs.~\ref{eqn:d_abs_approx} and
\ref{eqn:d_zpc_approx}, for isotropic features the required radiation
dose in x-ray microscopy increase as the fourth power of improvements
in spatial resolution.  Therefore in
Fig.~\ref{fig:xray_dose_versus_resolution} we show the required dose
assocated with $\snr$=5 x-ray imaging of protein features in 10
\micron\ thick amorphous ice as the size and thickness $t_{f}$ of the
features is changed, as calculated using the complete expression of
Eq.~\ref{eqn:theta_unified}.  As one decreases $t_{f}$ from 1 \micron\ 
towards 1 nm, the calculated dose changes by about 12 orders of
magnitude so we also show on this figure the threshold doses
associated with a variety of radiation-induced phenomena:
\begin{itemize}

\item The human LD$_{50}$ dose of about 4.5 Gray, which is the
  radiation dose that leads to
  to 50\%{} fatalities in human populations \cite{mole_bjr_1984}.

\item The incapacitation dose for rats and pigs, which is the dose at
  which these animals show an immediate cessation of normal function
  when the dose is administered over a time of minutes or
  less. Miniature pigs used in animal agriculture show immediate
  incapacitation upon exposure to gamma and neutron radiation doses of
  130 Gray \cite{chaput_radres_1970}, while male albino rats show
  immediate incapacitation at radiation doses of 153 Gray
  \cite{casarett_radres_1973}.

\item The LD$_{10}$ dose for the radiation-resistant bacterium
  \emph{Deinococcus radiodurans}.  This dose of 1.0--$2.5\times
  10^{3}$ Gray kills
  all but 10\%{} of these bacteria in sewage sludge or animal feed
  \cite{ito_abc_1983}.

\item The dose associated with immediate inactivation of muscle function.
  At an absorbed soft x-ray dose of about $2\times 10^{4}$ Gray, myofibrils
  will no longer contract in response to the addition of adeosine
  triphosphate or ATP \cite{bennett_jmic_1993}.

\item The dose associated with immediate structural changes in living
  cells. Chinese hamster ovarian cells show no immediate effect at soft
  x-ray doses of about $6\times 10^{2}$ Gray, but they show immediate
  and dramatic structural changes at a dose of $1.2 \times 10^{5}$
  Gray \cite{kirz_qrb_1995}.

\item The dose associated with changes in the carbon x-ray absorption
  near-edge structure (XANES) of polymers.  One can observe an
  $\exp[-1]$ decrease in the strength of the C=C absorption resonance in room
  temperature polymethyl methacrylate (PMMA) at a dose of about
  $1.2\times 10^{7}$ Gray \cite{zhang_jvstb_1995}, or in the C=O
  absorption resonance in PMMA at a temperature of 113 K at a dose of
  $1.3\times 10^{7}$ Gray \cite{beetz_jsr_2003}.

\item The so-called Henderson dose limit for fading of x-ray diffraction
  spots in cryo-cooled crystals of about $2\times 10^{7}$ Gray
  \cite{henderson_prslb_1990}.  It should be noted that a more recent evaluation has
  suggested that a dose of $3\times 10^{7}$ Gray represents a tolerable
  limit \cite{owen_pnas_2006}; in fact, there is some dependence on
  the length scale of bond-to-bond correlations (as measured from the
  diffraction angle of various Bragg peaks) as shown in Fig.~3 of
  \cite{howells_jesrp_2009}.  

\item The dose associated with mass loss in PMMA at a cryogenic
  temperature of 113 K.  Soft x-ray measurements have shown
  \cite{beetz_jsr_2003} an $\exp[-1]$ decrease in oxygen mass in PMMA
  films exposed to a dose of $6.0\times 10^{8}$ Gray at this
  temperature. It should be noted that PMMA is one of the more
  radiation-sensitive polymers, which is why it is used as a
  photoresist in electron beam lithography; polymers with aromatic
  rings as part of the monomer units are much more robust against mass
  loss.

\item The dose associated with ``bubbling'' of amorphous ice in
  electron microscopy \cite{dubochet_jmic_1982}, which involves the
  dissociation of H$_{2}$O to form hydrogen bubbles
  \cite{leapman_ultramic_1995}.  Though the dose varies with specimen
  conditions, an electron fluence of 5000 $e^{-}$/nm$^{2}$ at 100 kV
  (corresponding to a dose of about $1.6\times 10^{11}$ Gray) is
  representative.

\end{itemize}
Of these effects, all but the last two are measures either of
biological function, or specific atomic bonding or atom-to-atom
correlation distances.  For microscopy at a resolution of a few
nanometers, these effects do not affect the image (unless one is
utilizing XANES resonances for image contrast
\cite{ade_science_1992}); however, the last two measures (mass loss
and amorphous ice ``bubbling'') represent harder limits.  For this
reason, earlier studies on dose limits to resolution in x-ray
microscopy based on diffraction spot fading \cite{howells_jesrp_2009}
may be overly pessimistic.

\subsection{Ultimate thickness limits in x-ray microscopy}
\label{sec:xray_thickness_limits}

\begin{figure}[H]
%  \centerline{\includegraphics[width=0.9\textwidth]{xray_dose_energy_thickness_contour_100nm}}
  \centerline{\includegraphics[width=0.9\textwidth]{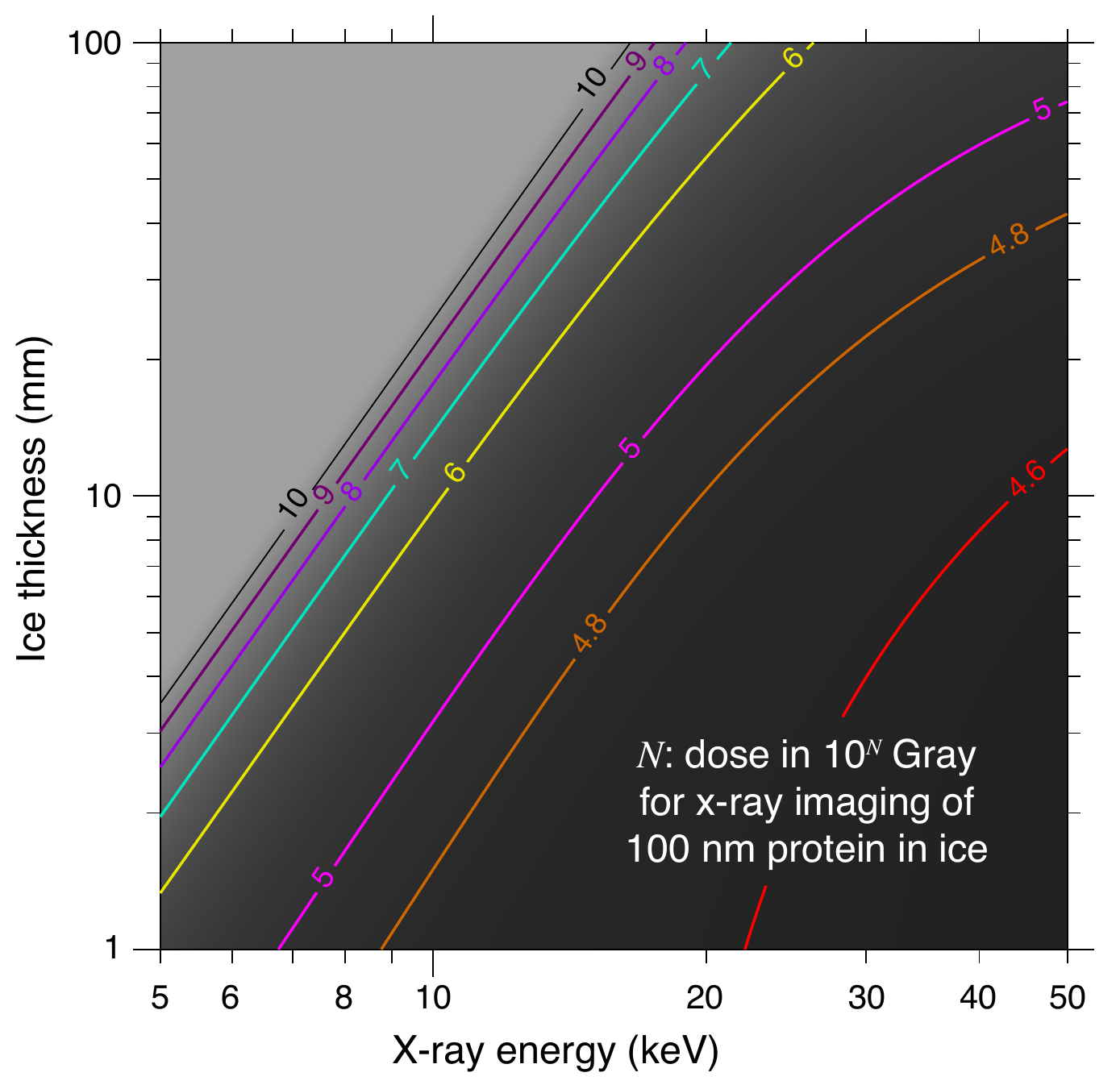}}
  \caption{\textcolor{blue}{Combined contour and brightness map of the required x-ray
    radiation dose in Gray for imaging 100 nm features in amorphous
    ice as a function of both x-ray photon energy and overall ice
    thickness.  This figure shows the lower of absorption or phase
    contrast imaging at each point; in nearly all cases phase contrast
    provides the lowest dose.  The grayscale image shows
    $\log_{10}({\rm Gray})$, with the overlaying contour line labeled
    6 representing a dose of $10^{6}$ Gray and so on.  Of course
    it would be very challenging to obtain amorphous ice over these
    organ-scale thicknesses, but on the other hand ice crystal
    artifacts that obscure features in few-nanometer-resolution cryo
    electron microscopy studies might be unnoticeable in 100 nm
    resolution imaging.}}
  \label{fig:xray_dose_energy_thickness_contour_100nm}
\end{figure}

In Sec.~\ref{sec:new_comparison}, we compared electron and x-ray
microscopy for 10 nm resolution imaging in amorphous ice thicknesses
up to 200 \micron\ and with x-ray energies as high as 20 keV, as shown
in Figs.~\ref{fig:electron_xray_dose_10nm} and
\ref{fig:xray_dose_energy_thickness_contour_10nm}.  We also saw in
Fig.~\ref{fig:xray_dose_versus_resolution} that the required x-ray
dose increases as the fourth power of improvements in spatial
resolution (corresponding to decreases in $t_{f}$).  We now consider a
somewhat different imaging regime: mesoscale resolution imaging with
$t_{f}=100$ nm for very thick specimens, such as might be desired for
x-ray imaging of whole mouse brains in neuroanatomical studies
\cite{hieber_scirep_2016,mizutani_scirep_2016}.  While these studies
are presently done with metal-stained, plastic-embedded specimens, we
have chosen to do a calculation for the case of protein in amorphous
ice for three reasons: 1) more direct comparison with the previous
figures; 2) because the general characteristics of inelastic and
plural elastic scattering shown in Figs.~\ref{fig:ice_x_cate} and
\ref{fig:epon_x_cate} are not qualitatively different between
amorphous ice and EPON; and 3) because it could be useful to consider
native image contrast should a means be found to minimize ice crystal
formation at 100 nm resolution in whole-organ-size specimens.  We
therefore show in
Fig.~\ref{fig:xray_dose_energy_thickness_contour_100nm} the results of
a calculation for amorphous ice thicknesses of 1 to 100 mm, and for
x-ray energies ranging from 5 to 50 keV.  As this figure shows, once
one has reached an x-ray energy sufficiently high enough to obtain
good penetration through the specimen, the required radition dose
shows very little variation with further changes in photon energy.
This is consistent with what one would expect from the thin-specimen
dose approximation of Eq.~\ref{eqn:d_zpc_approx2} even though the
background thickness is considerably higher here.  For real specimens
of this thickness, there will be significant problems in interpreting
2D images because of the overlap of features at many depth planes in a
projection, but this is what tomography untangles and again we would
expect dose fractionation \cite{hegerl_zn_1976} to mean that this
untangling does not come at a severe penalty in additional radiation
dose.

\section{Conclusion}

We have described a consistent approach to estimating the required
per-pixel quantum exposure $\bar{n}$ and associated radiation dose $D$
for both x-ray and electron microscopy of thick specimens, and have
applied these calculations to the example of imaging frozen hydrated
as well as plastic embedded biological specimens.  We have done so by
building upon previous work in x-ray
\cite{sayre_ultramic_1977,sayre_science_1977,golz_xrm1990,jacobsen_harder_xrm1990,jacobsen_xrm1996,schneider_ultramic_1998,wang_biotech_2013}
and electron
\cite{schroeder_jmic_1992,langmore_ultramic_1992,grimm_bj_1998}
microscopy, but with an accounting of inelastic and plural elastic
scattering, zero-loss energy filtering in electron microscopy, and
phase contrast in both x-ray and electron microscopy.  From a
radiation dose point of view, these calculations reinforce the idea
that electron microscopy offers advantages for specimens thinner than
about 1 micrometer (unless one wishes to obtain chemical contrast
using near-absorption-edge spectra
\cite{isaacson_optik_1978,rightor_jcpb_1997}, or if one seeks
maximum sensitivity in trace element mapping
\cite{kirz_nyacad306,kirz_nyacad342,kirz_sem_1980}), while x-ray
microscopy becomes the method of choice for thicker specimens.

\section*{Acknowledgement}

We gratefully acknowledge support from the National Institutes of
Health for support under grant U01 MH109100, and the Advanced Photon
Source, a US Department of Energy (DoE) Office of Science User
Facility operated under contract DE-AC02-06CH11357.  We also thank
Dr.~Qiaoling Jin for preparation of an EPON sample and Dr.~Kai He for
assistance with EELS measurements on that sample at Northwestern
University, and Dr.~Richard Leapman of the National Institutes of
Health for the amorphous ice EELS data shown in
Fig.~\ref{fig:ice_epon_eels}.

\end{document}